\documentclass[twocolumn]{article}

\usepackage{arxiv}

\usepackage[utf8]{inputenc} 
\usepackage[T1]{fontenc}    
\usepackage{hyperref}       
\usepackage{url}            
\usepackage{booktabs}       
\usepackage{amsfonts}       
\usepackage{nicefrac}       
\usepackage{microtype}      
\usepackage{color}
\usepackage{xcolor}

\usepackage[numbers]{natbib}
\usepackage{algorithm}
\usepackage{algorithmic}
\usepackage{rotating}
\usepackage{textcomp}
\usepackage{graphicx}
\usepackage{caption}
\usepackage{amsmath,bm}

\title {The Role of Regularization in Shaping Weight and Node Pruning Dependency and Dynamics} \begin{document}                     
\author{
  Yael Ben-Guigui \\
  The School of Electrical \\and Computer Engineering\\
  Ben-Gurion University\\
     \and
  Jacob Goldberger \\
  Engineering Faculty\\
  Bar-Ilan University\\
 
  \and
  Tammy Riklin-Raviv \\
  The School of Electrical \\and Computer Engineering\\
  Ben-Gurion University\\

}

\twocolumn[
  \begin{@twocolumnfalse} 
\maketitle
\begin{abstract}
The pressing need to reduce the capacity of deep neural networks has stimulated the development of network dilution methods and their analysis. While the ability of $L_1$ and $L_0$ regularization to encourage sparsity is often mentioned, $L_2$ regularization is seldom discussed in this context. 
We present a novel framework for {\bf weight} pruning by sampling from a probability function that favors the zeroing of smaller weights.
In addition, we examine the contribution of $L_1$ and $L_2$ regularization to the dynamics of {\bf node} pruning while optimizing for {\bf weight} pruning. We then demonstrate the effectiveness of the proposed stochastic framework when used together with a weight decay regularizer {on popular classification models} in removing 50\% of the nodes in an MLP for MNIST classification, 60\% of the filters in VGG-16 for CIFAR10 classification, {and on medical image models} in removing 60\% of the channels in a U-Net for instance segmentation {and 50\% of the channels in CNN model for COVID-19 detection.} For these node-pruned networks, we also present competitive weight pruning results 
that are only slightly less accurate than the original, dense networks.
\end{abstract}


  \end{@twocolumnfalse}
 ]

\section{Introduction}
The advent of deeper and wider neural networks
has made the need to reduce their storage and computational cost
critical, especially for certain real-time applications. 
In particular, features such as model size, memory footprint, the number of floating point operations (FLOPs) and power usage which are directly related to the number of network parameters must be considered in resource-constrained setups such as mobile phones and wearable devices. 
In recent years there has been a significant effort to address the apparent trade-off between inference accuracy and reliability which requires sufficient network capacity and parameters' frugality, see~\cite{blalock2020state} and references therein. However, having adequate network capacity and maintaining network compactness are not necessarily contradictory goals.  
It has been shown that neural networks are usually overparametrized and therefore can be pruned without hardly loss in accuracy \cite{han2015deep, ullrich2017soft, molchanov2017variational}. Moreover, overparameterized neural networks can easily overfit and even memorize random
patterns in the data ~\cite{zhang2016understanding}, and thus lack the ability to generalize at inference time.

%
One straightforward approach to reducing the number of neural network' parameters consists of systematic weight or node pruning while training. In~\cite{han2015learning} the pruning phase was applied to parameters of an already trained and optimized neural network in an attempt to maintain similar accuracy. Most current methods follow this strategy since pruning from scratch was shown to be inferior \cite{zhang2016understanding,louizos2017bayesian}. 

Since random pruning may reduce network accuracy~\cite{blalock2020state},
a pruning decision is usually based on some scoring criterion, e.g., absolute values or contributions to network activation or gradients.
In this context it is worth noting that the optimal pruning criterion is debatable. In~\cite{hinton1986learning} it was suggested that a weight's magnitude is indicative of its usefulness in reducing the error, while other approaches have suggested more sophisticated "importance" criteria, and utilized the Hessian of the loss function~\cite{lecun1990optimal,hassibi1993second} or a Taylor expansion to approximate its changes due to pruning~\cite{molchanov2016pruning}. However, the latter approaches involve additional computational cost during training.     
The other main source of variability across different pruning algorithms are the timing and the way the pruning is carried out.
Although setting thresholds below which nodes or edges are removed is useful ~\cite{han2015learning} a stochastic approach that defines the probability for pruning is more elegant. However, some stochastic pruning algorithms involve thresholding as well, either because the weight zeroing is done temporarily - as in  Dropout based approaches~\cite{gomez2019learning,molchanov2017variational} or because the probabilistic gating
mechanism is not binary~\cite{zhao2019variational}.
Obviously, setting a threshold as well as the timing and frequency of its application requires determining the right balance between the pruning ratio and accuracy. All these hyper-parameters are network-dependent and if the threshold is applied as part of a post-process, it cannot be learned during training.   

A network can be pruned at different levels. While weight pruning is highly  prevalent \cite{han2015deep, ullrich2017soft, molchanov2017variational}, node pruning, i.e., the removal of a node along with all of its in-going and out-going connections is considered to be more effective, reduces computational cost and is more suitable for hardware and software optimization \cite{augasta2013pruning, wen2016learning, louizos2017bayesian, neklyudov2017structured}.  
In the same manner, pruning can be applied to redundant channels~\cite{he2017channel} and kernels or filters~\cite{li2016pruning,luo2017thinet,huang2018learning,he2019filter} to simplify convolutional neural networks or to directly reduce the FLOPS ~\cite{aflalo2020knapsack}. 
The level of pruning can be changed in any stage of the optimization, depending on the pruning objective~\cite{pasandi2020modeling}. However, to the best of our knowledge, simultaneous pruning of different levels with a single pruning objective has never been addressed. Moreover, most weight pruning algorithms achieve poor node pruning results, especially when weight pruning is done at the end of a training process based on some predefined criterion, e.g., the probability that all {weights} connected to a particular node will be zeroed is exponentially low. 

In a recent line of works, regularization was employed to encourage sparsity.
In ~\cite{louizos2017learning} a differentiable approximation to $L_0$ regularization via the hard concrete distribution was used.  
Other methods have utilized the $L_1$ norm as an alternative approximation to $L_0$, 
e.g.,~\cite{liu2017learning,huang2018data}. However, while $L_2$ regularization is occasionally incorporated as part of a network's loss function \cite{han2015learning, han2015deep}, its effect on the dynamics of the sparsity-constrained training process has not been examined.

The key contribution of the proposed study lies in  utilizing weight decay regularization to facilitate pruning and understanding its role in manipulating pruning dynamics. Rather than using a fixed pruning criterion, we define a probability function and a gating mechanism such that weights with lower absolute values are more likely to be removed. Regularization then plays a triple role. Beyond its clear advantage in reducing overfitting, it decreases weight' values, thus increasing their pruning probability. Moreover, it turns out that while optimizing for weight pruning, node pruning is increased as well. Our experiments show that an application of a weight regularizer suppresses the tendency of edges to compensate for pruned edges that are connected to the same nodes. 

The proposed method, dubbed Weight to Node Pruning (WtoNP), is applied to several standard network architectures, for both classification and segmentation and is compared to state-of-the-art weight and node pruning algorithms. Specifically we demonstrate its strength using a multilayer perceptron (MLP) and a VGG-16 for MNIST and CIFAR-10 classification (respectively) and the U-Net for segmentation of cell microscopy data. The results show that our method is on a par with weight pruning algorithms and with node pruning algorithms. Moreover, the pruning of both weights and nodes has the clear advantage of reducing higher ratios of the parameters, thus saving storage space while more effectively reducing the computational cost. 
The code of the proposed pruning method, which can be easily adapted to neural networks of different architectures, will be made available upon the acceptance of this manuscript.


\section{Stochastic weight pruning and regularization}
\label{sec:stochastically}
Given a network architecture and a training dataset we aim to learn a sparse parameter set $\theta$ by optimizing a suitable objective function. Our explicit goal is weight pruning 
but we show that if a weight regularization function is added to the cost function, pruning of the network nodes also takes place. 
Our pruning method is based on zeroing each edge weight in a stochastic manner where the probability of a network parameter to be retained is proportional to its magnitude.

Let $h(\cdot,\theta)$ define a non-linear function parametrized by $\theta$ that represents a neural network. 
%
We define the network's loss function that includes the regularisation terms $L(\cdot)$ with respect to a given set of input-output pairs $\{x_i,y_i\}_{i=1}^n$ and a set of weights $\theta=(w_1,...,w_m)$ as follows:
\begin{equation}
    L(\theta) = \frac{1}{n}\sum_{i=1}^n 
     l(h(x_i;\theta),y_i) + \lambda \mathcal{R}(\theta) 
\end{equation}
where $\mathcal{R}(\theta) = ||\theta||^p_p$ is the  $L_p$ norm regularization function of the weights, $\lambda$ is a non-negative hyperparameter
and $l(\cdot)$ corresponds to the model loss function. 
In \cite{louizos2017learning} $L_0$ regularization was suggested to encourage weight sparsity. Here, in contrast, we focus on either $L_1$ or $L_2$ regularization. While the  $L_1$ norm is often used as a differentiable approximation of $L_0$,  $L_2$ norm which encourages weight decay~\cite{hanson1989comparing,hinton1986learning} has not been used explicitly to encourage network sparsity.
In the following, we show that once a stochastic weight gating function is applied that favors zeroing weights with smaller magnitudes, $L_2$ and $L_1$ regularizations are advantageous. 

Recall that for $L_2$, i.e., when $\mathcal{R}(\theta) \equiv ||\theta||^2_2 = \sum\limits_{j=1}^m {w_j^2}$ 
the update step of a weight $w_j$ {by gradient decent optimizer} is the following:
\begin{align}
{w}_j^t = (1-\lambda') w_j^{t-1} - \epsilon (\nabla_{w_j}l(\theta^{t-1}))
\end{align}
where $t$ counts the update steps, $\epsilon$ is the learning rate and $\lambda' = \lambda\epsilon  $.
Following~\cite{hanson1989comparing,hinton1986learning} we set $0 <\lambda' < 1.$ When the gradient of $l(\theta)$ with respect to $w_j$ is negligible, $w_j$ shrinks in each step by a factor of  $1-\lambda'$.  \\
Similarly, the update step in the presence of $L_1$ regularization is as follows:
\begin{align}
{w}_j^t =  w_j^{t-1} - \lambda'~\mbox{\bf sign}(w_j^{t-1}) -  \epsilon(\nabla_{w_j}l(\theta^{t-1}))
\end{align}
which means that whether positive or negative, if $\nabla_{w_j}l(\theta^{t-1})$ is negligible, $w_j$ is ``pushed" to zero by $\lambda'$ in each gradient descent step.

Our goal is to stochastically prune weights based on their magnitude, where weights with smaller magnitudes have a higher probability to be pruned.
Each weight can be zero with a different probability, according to Bernoulli distribution ($\sim Bern(\cdot)$), depends on the weight magnitude.
Therefore we use a function for mapping the weight magnitude to a probability. Let $\varphi: [0, \infty) \rightarrow[0, 1]$ denote a monotonically increasing function and let $z_j$ denote a binary gating variable $z_j \sim Bern\left(\varphi (|w_j|)\right).$
We propose a modified weight update during the parameter learning process as follows:
\begin{align}
\tilde{w}_j^{t}&  = w_j^{t-1} - g\left(\nabla_{w_j}L(\theta^{t-1})\right) \hspace{1cm} j=1,...,m \\
z_j^t & \sim Bern\left(\varphi (|\tilde{w}_j^t|)\right) \\
w_j^t  & = \tilde{w}_j^t \cdot z_j^t 
\end{align}
where $g()$ is an optimizer scheme.
Several even functions can be considered to represent $\varphi$, e.g., a Gaussian shaped  function $1-\exp( -\frac{1}{2}aw^2)$, We can also use
$1-4\sigma(aw)(1-\sigma(aw))$ such that $\sigma()$ is the sigmoid function and $a$ is a parameter that controls its slope.

\begin{figure}[ht]
\centering
\begin{tabular}{l}
\includegraphics[width=1\linewidth]{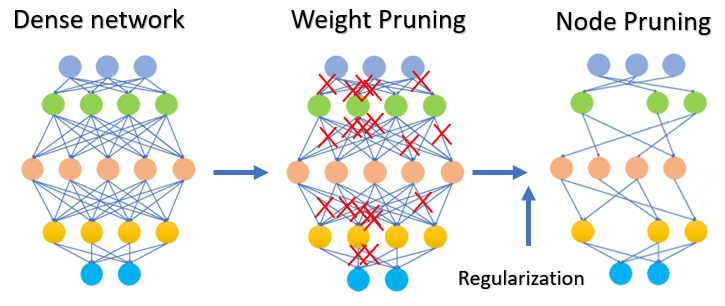}
\end{tabular}
\caption{{\bf Illustration of the WtoNP algorithm main concept.} The WtoNP algorithm is applied to a trained network by sampling weights and pruning them based on their magnitude. If a weight regularization function is added to the cost function, pruning of the network nodes also takes place. }
\label{fig:GAbstract}
\end{figure}

The Weight to Node Pruning (WtoNP)
is summarized in~ Algorithm~\ref{alg:zeroed network weights} and the main concept is illustrated in figure ~\ref{fig:GAbstract}. It is applied to a trained network performing an additional training session that enforces sparse solutions by stochastically selecting a set of weights to be dropped.
The proposed strategy of random zeroing of  network components resembles the Dropout algorithm \cite{dropout2014}. The main difference is that dropout
temporarily eliminates different fractions of the weights in each forward-backward session of a single instance and then restores the original weights' values. In our approach when a weight $w$ is zeroed its previous value is discarded. In \cite{gomez2019learning}, on the other hand, the Dropout is applied to reduce  model dependency on the unimportant edges, thereby potentially reducing the performance degradation as a result of post-processing pruning (via threshold application). In contrast, the proposed joint edge pruning and model optimization explicitly targets both the performance and the sparsity of the learned network.
{Louizos et al. use an adaptation of concrete dropout and Molchanov et al.~\cite{molchanov2017variational} use variational dropout, both with trainable drop rates to the weights of the network. In these two works, in contrast to our method the number of trainable parameters was doubled.}

In general, there is no direct connection between weight pruning and node pruning and there must be a different explicit loss function for each of the two pruning goals. In other words, weight pruning algorithms do not encourage other edges that enter the same node (a.k.a. neighbouring edges) to be pruned together. In our scheme, a zeroed edge can move away from zero but since the updated weight is still small there is a good chance that it will be zeroed again and so on.
In that case we can expect that the weights of the neighboring edges will become larger to compensate for the removed edge.
However, if the network parameters are regularized, the rate of growth of edges that are neighbors of the zeroed edge is suppressed. This in turn causes the entire node to be less informative for the global network decision such that gradually the magnitudes of its edges will decrease as well until finally they will be zeroed. We demonstrate this dynamics for various network architectures in the experiment section.

\begin{algorithm}[tb]
  \caption{Weight to Node Pruning (WtoNP) Algorithm}
  \label{alg:zeroed network weights}
\begin{algorithmic}
\STATE {\bfseries Training dataset:} input-output pairs $(\mathbf{x},\mathbf{y}) \equiv \{x_i,y_i\}_{i=1}^n$ \\
\STATE {\bfseries Parameters:}  Trainable network weights 
${\bf\theta}=(w_1,...,w_m)$
\STATE {\bfseries Trained neural network:} $h(\cdot,\theta)$ with respect to some loss function $l({\bf y}, h({\bf x},\theta))$ and some optimizer scheme $g(\cdot)$
\STATE
  \FOR{each minibatch:}
\STATE Compute  gradients with respect to the loss:\\
 $  L(\theta) = \frac{1}{n}\sum_{i=1}^n 
     l(h(x_i;\theta),y_i) + \lambda \|\theta\|_p,~ p\in\{1,2\} $.
 \STATE Update network parameters:
  \FOR{j = 1 to m}
  \STATE $w_j \leftarrow w_j- g\left(\nabla_{w_j}L(\theta)\right)$
  \ENDFOR
\STATE Stochastically prune weights based on their magnitude: 
     \FOR{j = 1 to m}
      \STATE $Z_j \sim Bern(\varphi(|w_j|))$
      \STATE set $w_j \leftarrow w_j \cdot z_j$
      \ENDFOR
\ENDFOR 
\end{algorithmic}
\end{algorithm}

\section{Pruning performances, comparisons and analysis}
To validate the effectiveness of our method, we applied it to two different classification tasks and instance segmentation.
We employed several regularization terms: $L_1$, $L_2$ and elastic-net, as well as no regularization. 
For all tasks, we initially reached a baseline model by training a full-size network until convergence. We then applied our pruning strategy to the trained network by performing a combined training and pruning session. 


\subsection{Implementation details}
We tested the proposed pruning algorithm on four different network architectures performing four different tasks. 
{The networks were trained using Adam optimizer~\cite{kingma2014adam} with the following (default) hyper-parameters: learning rate=0.001, $\beta_1$=0.9 and $\beta_2$=0.999. The only exception was the learning rate of the VGG-like network which was set to $0.0005$.
The proposed pruning technique is implemented as a Keras callback or fastai callback and therefore can easily be incorporated in the training of any network regardless of its architecture. We define the stochastic function $\varphi$ as follows:
\begin{equation}
    \varphi(w) = 1 - 4\sigma(aw)(1-\sigma(aw)),
\end{equation}
where $\sigma$ denotes the standard sigmoid function. The hyperparameter $a$ which controls the slope of the sigmoid, was tuned using the validation set.
In our experiments $a$ took values in the range of $[1e-2, 1e-5].$}\\

{\bf MLP-300-100 for MNIST classification:}
The MLP-300-100~\cite{lecun1990optimal} is a fully connected network with two hidden layers, with 300 and 100 neurons each with 267K parameters.
The baselines were trained for 200 epochs and the networks were diluted for another 200 epochs.
We set $\lambda=1e-4$ for all regularizations terms in the pruning sessions.

{\bf VGG-like for CIFAR-10 classification:}
The VGG-like network was inspired by the original VGG-16 ~\cite{simonyan2014very}. It is a standard
convolution neural network with 13 convolutional layers followed by two fully connected layers with 512 and 10 neurons,
respectively. The total number of trainable parameters is 15M.
We used the CIFAR-10 dataset ~\cite{krizhevsky2009learning} to evaluate this model. 
The network was trained for 250 epochs with $L2$ regularization. For the pruning session we used $\lambda=5e-5$ and $\lambda=1e-5$ for $L_2$ and $L_1$ regularization, respectively.

\begin{figure}[ht]
\centering
\begin{tabular}{l}
~~~~Cell image ~~~~~~~~~~~~Manual seg.~~~~~~~~~~~Predicted seg.\\
\hspace{-1.2cm}
\includegraphics[trim={1cm 1.4cm 1cm 0.6cm},clip,width=1.24\linewidth]{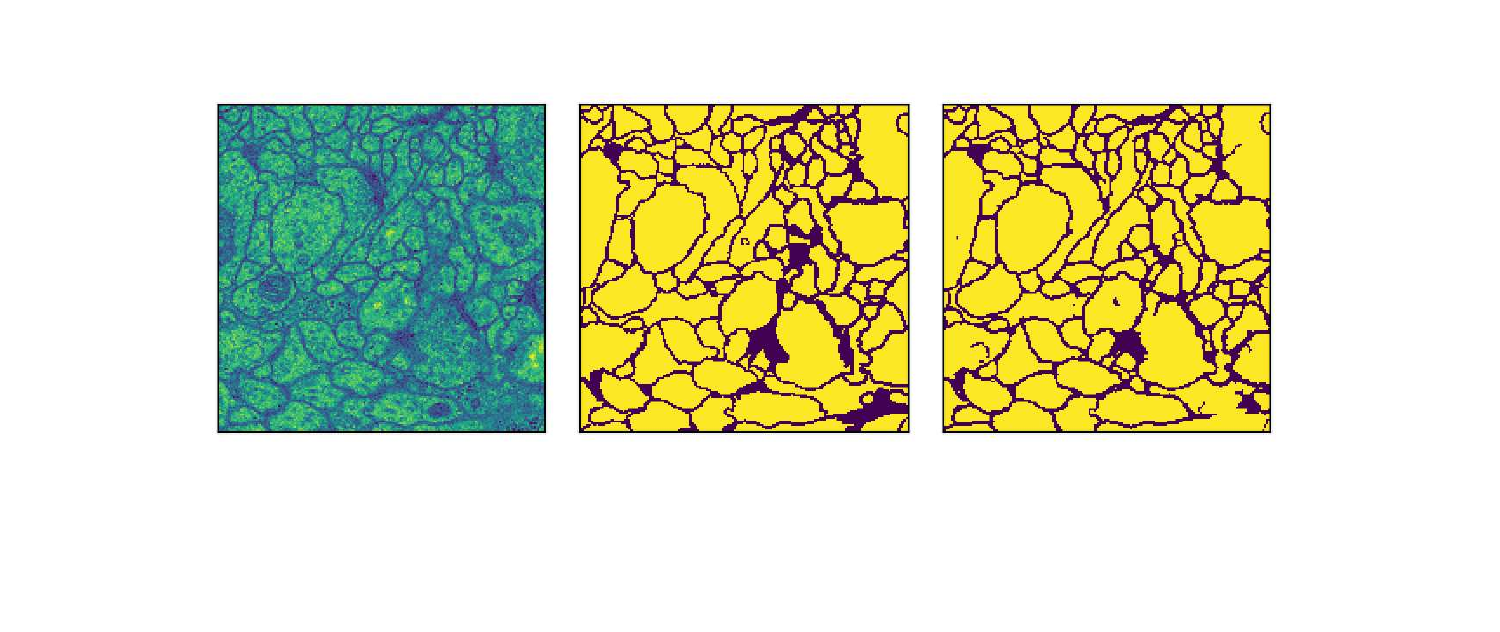}
\end{tabular}
\caption{{\bf A microscopy image along with its segmentations.} A cell microscopy image (left) along with its manual segmentation (middle) and its segmentation prediction by the U-Net following pruning of 95\% of its weights.}
\label{fig:CellMicroscopyImage}
\end{figure}

{\bf U-Net for cell segmentation in microscopy images:}
The U-Net architecture which was initially introduced in~\cite{ronneberger2015u} is a symmetrical fully convolutional network with skip connections, comprising a contracting path and an expansive path. 
The tested cell segmentation task and the dataset used are part of the ISBI 2012 segmentation challenge\footnote{http://brainiac2.mit.edu/isbi\_challenge/}.
For this experiment we used the serial section Transmission Electron Microscopy (ssTEM) dataset of the Drosophila first instar larva ventral nerve cord (VNC)~\cite{arganda2015crowdsourcing}. 
{An example of a microscopy image from the tested dataset is presented in figure~\ref{fig:CellMicroscopyImage}.
To demonstrate the quality of the segmentation results, a manual segmentation (middle) of the cell image on the left is presented along with the U-Net prediction (right) after the pruning of 95\% of its weights.}
The training data include 30 images (512 $\times$ 512 pixels) randomly partitioned into patches (128$\times$128 pixels), where each training example is paired with a manual (ground truth) segmentation. For the test dataset which contains 30 images as well, evaluation was performed by the challenge organizers since the ground truth annotations were confidential. 
The U-Net was trained for 3000 epochs each session using cross entropy loss and data augmentation as in~\cite{ronneberger2015u}.
For the pruning sessions, we set the coefficient of all regularizations terms to $\lambda=1e-4.$

\begin{figure}[ht]
\centering
\begin{tabular}{ccc}
COVID-19 &Pneumonia &No-Findings.\\
\hspace{-0.2cm}
\includegraphics[width=0.27\linewidth]{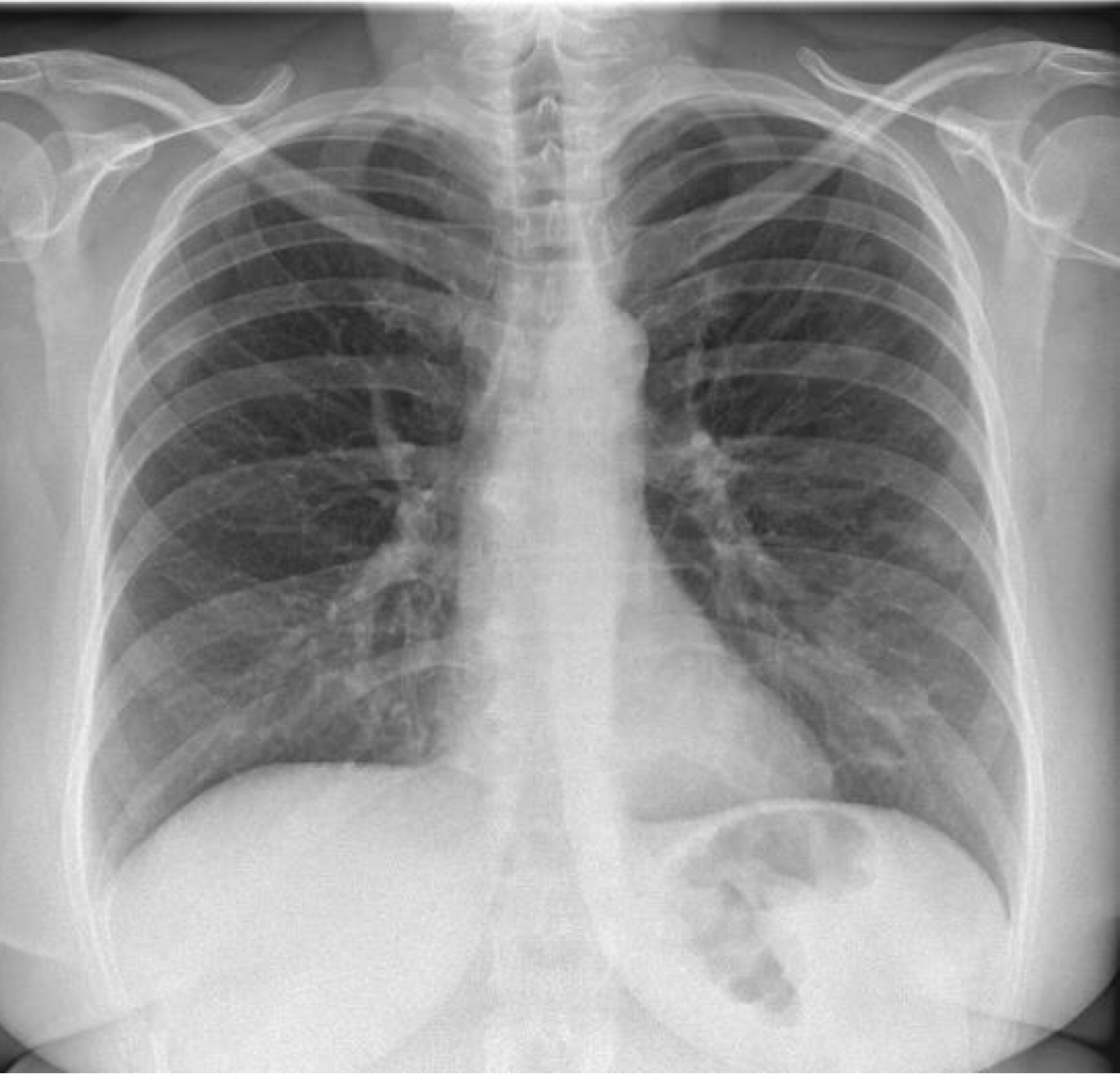}&
\includegraphics[width=0.27\linewidth]{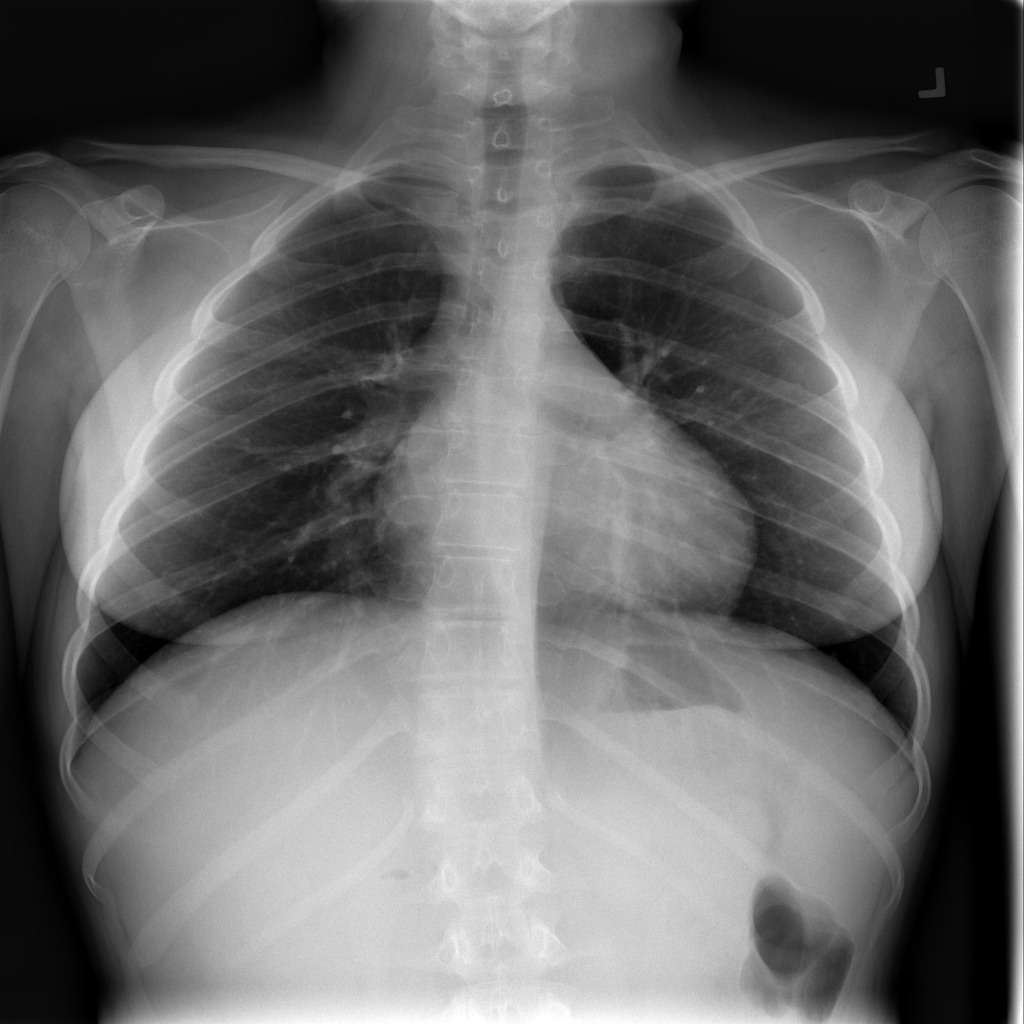} &
\includegraphics[width=0.27\linewidth]{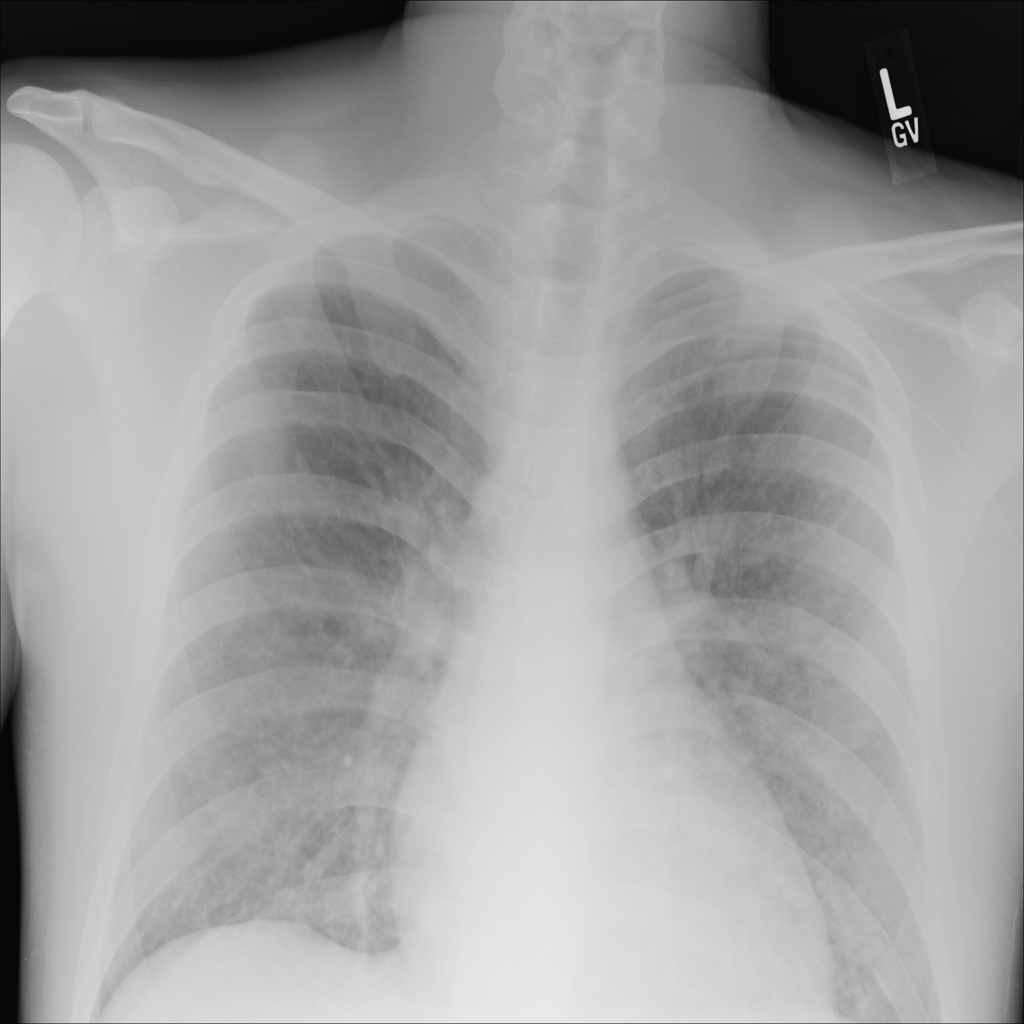} 
\end{tabular}
\caption{Chest X-ray images of COVID-19 case (left) Pneumonia case (middle) and  No-Findings.}
\label{fig:CovidImage}
\end{figure}

{\bf DarkCovidNet Model for automatic COVID-19 detection using raw chest X-ray images:}
The DarkCovidNet architecture introduced in~\cite{ozturk2020automated} and inspired by the DarkNet architecture. This model contains 17 convolution layers and uses for multi-class classification (COVID vs. No-Findings vs. Pneumonia).
We used the dataset with which the model was trained in the original paper. This dataset contains 125 X-ray images diagnosed with COVID-19, 500 no-findings and 500 pneumonia class frontal chest X-ray images from two different sources~\cite{cohen2020covid,wang2017hospital}. Figure ~\ref{fig:CovidImage} present example for chest X-ray images with COVID-19 case (left) Pneumonia case (middle) and  No-Findings.
80\% of the X-ray images are used for training and 20\% for test.
The model has been trained using the code published by ~\cite{ozturk2020automated} for 100 epochs.
For the pruning sessions, we used $\lambda=5e-4$ and $\lambda=1e-4$ for $L_2$ and $L_1$ regularization, respectively.


\begin{table*}[t]
\begin{center}
\caption{Comparison to weight pruning algorithms}
\label{table:WeightsResults}
\small
\centering
\begin{tabular}{ c|c|l|l|l }
  \hline
 \scriptsize{Model}
& \multicolumn{2}{|c|}{\scriptsize{Method}}
& \scriptsize{Error rate[\%] $\downarrow$ }
& \scriptsize{\% Pruned Weights $\uparrow$ }
\\
 \hline

MLP-300-100&\multicolumn{2}{|l|}{LWC~\cite{han2015learning}}& $1.64 \rightarrow 1.58$  & $91.77$ \\

(MNIST) &\multicolumn{2}{|l|}{L-OBS~\cite{dong2017learning}}& $1.76 \rightarrow 1.82$  & $93.00$ \\

&\multicolumn{2}{|l|}{Zhang et al~\cite{zhang2018systematic}}& $1.60 \rightarrow 1.60$ & $95.63$ \\

& \multicolumn{2}{|l|}{SWS~\cite{ullrich2017soft}}& $1.89 \rightarrow 1.94$ & $95.70$ \\

&\multicolumn{2}{|l|}{DNS~\cite{guo2016dynamic}} & $2.28 \rightarrow 1.99$  & $98.21$ \\

&\multicolumn{2}{|l|}{GSM~\cite{ding2019global}} & $1.81 \rightarrow 1.82$ & $98.34$ \\

&\multicolumn{2}{|l|}{Autoprune~\cite{xiao2019autoprune}} & $1.72 \rightarrow 1.78$ & $98.75$ \\

&\multicolumn{2}{|l|}{WtoNP} & $1.53 \rightarrow 1.71$ & $98.45$ \\

 \hline
VGG-like&  \multicolumn{2}{|l|}{BLIND~\cite{salama2019prune}} & ${6.75 \rightarrow 6.59}$   & $86.10$ \\

(CIFAR-10)&\multicolumn{2}{|l|}{Zhu et al~\cite{zhu2018improving}}& $6.49 \rightarrow 6.69$ & $88.24$\\

& \multicolumn{2}{|l|}{Huang et al~\cite{huang2018learning}} & $7.23 \rightarrow 10.63$ & $92.80$ \\

& \multicolumn{2}{|l|}{DCP~\cite{zhuang2018discrimination}}& $6.01 \rightarrow 5.43 $ & $93.58$ \\

&\multicolumn{2}{|l|}{SparseVD~\cite{molchanov2017variational}  } & $7.55\rightarrow7.55$  & $98.46$\\
& \multicolumn{2}{|l|}{WtoNP} & $6.46 \rightarrow 6.69$ & $93.86$ \\
\hline
\end{tabular}
\end{center}
\end{table*}

\begin{table*}[ht]
\begin{center}
\caption{Comparison to node pruning methods}
\label{table:NodesResults}
\small
\centering
\begin{tabular}{ c|l|l|l|l|l }
  \hline
 \scriptsize{Model}
& \multicolumn{2}{|c|}{\scriptsize{Method}}
& \scriptsize{Error rate[\%] $\downarrow$ }
& \scriptsize{\% Pruned Nodes $\uparrow$ }
& \scriptsize{\% Pruned Weights $\uparrow$  }
\\
 \hline
MLP-300-100&\multicolumn{2}{|l|}{ SparseVD~\cite{molchanov2017variational}}&$1.60\rightarrow1.80$  & $41.05$  & $97.80$\\

(MNIST)&\multicolumn{2}{|l|}{ BC~\cite{louizos2017bayesian}  } & $1.60\rightarrow1.80$  & $67.14$  & $89.20$\\

&\multicolumn{2}{|l|}{ $L_0$~\cite{louizos2017learning}}& $1.60\rightarrow1.80$  & $67.30$ & $-$ \\

&\multicolumn{2}{|l|}{NeST ~\cite{dai2019nest}}       & $1.29$  & $16.13*$  &$97.05$\\

&\multicolumn{2}{|l|}{ Autoprune~\cite{xiao2019autoprune}}& $1.60\rightarrow1.82$  & $69.01$  &$-$\\

&\multicolumn{2}{|l|}{WtoNP}                                 & $1.40 \rightarrow 1.73$ & $49.00$  & $98.30$\\

\hline
VGG-like&  \multicolumn{2}{|l|}{Li et al~\cite{li2016pruning} }&   $6.75 \rightarrow 6.60$    & $37.12$  & $ 64.00$\\

(CIFAR-10)&   \multicolumn{2}{|l|}{StructuredBP~\cite{neklyudov2017structured}}&    $ 7.20\rightarrow7.50 $  & $  75.17$& $-$ \\

& \multicolumn{2}{|l|}{Liu et al~\cite{liu2019channel}}&   $7.53 \rightarrow 8.25$    & $47.35$  & $ 36.00$\\

& \multicolumn{2}{|l|}{VCNN~\cite{zhao2019variational}}  &    $ 6.75\rightarrow6.82  $  & $  62.00$& $73.34 $ \\

& \multicolumn{2}{|l|}{Wang et al~\cite{wang2019reliable}}&    $ 6.87\rightarrow6.85  $  & $ 59.10$& $91.80 $ \\

&\multicolumn{2}{|l|}{WtoNP} & $6.46 \rightarrow 7.71$ & $60.34$  & $83.53$\\

\hline
\end{tabular}

\begin{tabular}{l|l|l|l|l|l|l }
\multicolumn{7}{c}{}\\
\multicolumn{7}{c}{Medical image networks pruned with WtoNP}\\
\hline
\scriptsize{Model} & \multicolumn{2}{|c|}{ \scriptsize{segmentation score (rand)$\uparrow$}}& \multicolumn{2}{|c|}{\scriptsize{segmentation score (MI)$\uparrow$}}& \scriptsize{\% Pruned Nodes $\uparrow$ }& \scriptsize{\% Pruned Weights $\uparrow$}
\\
\hline

U-Net  & \multicolumn{2}{|l|}{$96.37\rightarrow96.43$}& \multicolumn{2}{|l|}{$98.20\rightarrow98.22$} & $60.60$& $96.94$\\
\hline

\scriptsize{Model$\uparrow$}&
\scriptsize{precision$\uparrow$}& \scriptsize{recall$\uparrow$}& \scriptsize{f1-score$\uparrow$}& \scriptsize{accuracy$\uparrow$}& 
\scriptsize{\% Pruned Nodes$\uparrow$ }& 
\scriptsize{\% Pruned Weights$\uparrow$}\\
\hline

DarkCovidNet& $89\rightarrow89$& $83\rightarrow82$ & $85\rightarrow85$ & $84\rightarrow85$& $48$ & $97$\\
\hline

\end{tabular}
\end{center}
\end{table*}

\begin{table}[bth]
\begin{center}
\caption{Pruning 90\% of VGG-like weights on CIFAR-10}
\label{table:structeredVGG}
\small
\centering
\begin{tabular}{ c|l|c|c }
  \hline
\scriptsize{Regularization}
&\scriptsize{Method}
& \scriptsize{Zero node Ratio$\uparrow$}
& \scriptsize{Error rate[\%]$\downarrow$}
\\
 \hline
$L_2$ &WtoNP  &  $26.9\pm3.0$ & $\bf{9.0\pm0.2}$\\

& SKeras\cite{zhu2017prune}  &  $\bf{30.3\pm5.8}$ & $10.4\pm0.6$\\

& TD\cite{gomez2019learning}  &  $0$ & $9.6\pm0.4$\\
\hline

$L_1$&WtoNP &   $\bf{64.1\pm2.3}$ & $9.8\pm0.1$\\

& SKeras\cite{zhu2017prune} &  $53.0\pm8.9$ & {$\bf8.3\pm0.6$}\\

&TD \cite{gomez2019learning}  &  $0$ & $10.0\pm0.2$\\

\hline
without &WtoNP&  $\bf{1.7\pm0.5} $ & $\bf{8.2\pm0.3}$\\

& SKeras\cite{zhu2017prune}  &  $ 0.1\pm 0.1 $ & $17.7\pm0.8$\\

& TD \cite{gomez2019learning} &  $0$ & $10.5\pm0.3$\\

\hline
\end{tabular}
\end{center}
\end{table}


\subsection{Comparison of the different pruning techniques}
We compared the WtoNP to the state-of-the-art weight and node pruning methods as shown in
Table~\ref{table:WeightsResults} and Table~\ref{table:NodesResults}, respectively.
The methods were evaluated based on their classification errors before and after the pruning procedure and the percentages of pruned weights or nodes.
For MLP-300-100, we calculated the node percentage including the input layer, marking with~* pruning methods that did not consider it. For the MLP-300-100 we obtained a weight pruning percentage of more than 98\%, while maintaining higher accuracy than other pruning methods with similar pruning ratios.
For the VGG comparison, `pruned nodes' indicate the kernels' prune percentage. Note the relatively low error rate (6.69) with the high percentage of weight pruning (93.86\%) that was obtained using our WtoNP. In addition, our method presented competitive nodes pruning performances, by pruning 60\% of the VGG network kernels. 
{Pruning results of networks for medical application are barely reported in the literature}.
Bear in mind that though less than 4\% of the U-Net weights remained, the segmentation scores were improved.{For the DarkCovidNet Model we obtained 97\% weight pruning similar classification reports, with higher accuracy and minor drop on the recall.} 
The results reported in Tables~\ref{table:WeightsResults}-\ref{table:NodesResults} were obtained using $L_2$ regularization.    

We next tested the impact of $L_1$ and $L_2$ regularizations on pruning performance for the VGG-like network using the WtoNP and two other pruning approaches that originally did not include a regularization term. Specifically, we used a standard Keras implementation of magnitude-based weight pruning~\cite{zhu2017prune} and an implementation of the targeted Dropout algorithm~\cite{gomez2019learning} with a TensorFlow backend \footnote{https://pypi.org/project/keras-targeted-dropout/}.  Table~\ref{table:structeredVGG} presents the comparisons for pruning 90\% of the weights. It shows that the presence of a regularization term facilitated node pruning for the magnitude-based weight pruning method in~\cite{zhu2017prune} as well.
On the other hand, regularization did not influence the targeted Dropout pruning method since the pruning in that case is done as a post-processing step. 

\subsection{Analysis of pruning structure for different regularizations}
{In this section, we presents further pruning results obtained using $L_1, L_2$ and elastic net regularizations for the VGG, MLP, U-Net and DaekCovidNet networks and investigate the pruning structure in particular layers for the MLP-300-100 and the U-Net architectures.}

{Table~\ref{tbl:MLPResults} presents further pruning results obtained using $L_1, L_2$ and the elastic-net regularizations with respect to no regularization at all for the different neural network architectures. For fair comparison, in all experiments we kept the percentage of pruned weights fixed and present the respective accuracy and percentage of pruned nodes. As expected, there is a trade-off between these measures, where $L_1$ is preferable for the VGG-like network when the main objective is to maximize the percentage of pruned nodes, whereas $L_2$ works the best in all respects for the U-Net. We stress that in all experiments the pruning process was applied directly to the weights and not to the nodes (kernels), while tuning $\lambda$ and $a$ to achieve the desired percentage of pruned weights.}

\begin{table*}[h!]
\begin{center}
\caption{Pruning Results for Different Regularizations}
\label{tbl:MLPResults}
\small

\begin{tabular}{ c|c|l|l|l|c }
  \hline
 \scriptsize{Model}
& \scriptsize{Reg}
& \multicolumn{2}{|c|}{\scriptsize{Top-1 Error[\%]}}
& \scriptsize{\% Pruned nodes }
&\scriptsize{\% Pruned weights}
\\
    \hline
    MLP-300-100 &  $L_1$ &\multicolumn{2}{|l|} {$2.57\pm0.2$ }& $\bm{42.30\pm0.8}$ & 92   \\
    &$L_2$ &\multicolumn{2}{|l|}{$ \bm{ 1.66\pm0.1} $}   & $  34.80\pm0.7  $ &\\
    &{Elastic-net }&\multicolumn{2}{|l|}{$ 2.28\pm0.2 $} &  $ 41.40\pm0.7$ &\\
    & No-regularization &\multicolumn{2}{|l|}{$ 1.71\pm0.1$} & $ 13.30\pm0.4 $  &\\
    \cline{2-6}
    &  $L_1$ &\multicolumn{2}{|l|}{$2.28\pm0.1$} & ${45.90\pm0.7}$ & 97   \\
    &$L_2$ &\multicolumn{2}{|l|}{ $ \bm{ 1.76\pm0.1} $}  & $  \bm{47.40\pm0.7}  $ &\\
    &{Elastic-net }&\multicolumn{2}{|l|}{$ 1.98\pm0.1 $} &  $ 45.80\pm0.7$ &\\
    & No-regularization &\multicolumn{2}{|l|}{$ 2.18\pm0.3$}   & $ 16.07\pm2.2 $  &\\
    \hline
    VGG-like &  $L_1$ &\multicolumn{2}{|l|}{$9.79\pm0.1$ }& $\bm{ 64.07\pm2.3}$ & 90   \\
    &$L_2$ &\multicolumn{2}{|l|}{$ 8.97\pm0.2$}  & $26.96\pm3.1 $  &\\
    & No-regularization &\multicolumn{2}{|l|}{$\bm{8.17\pm0.3}$}  & $1.68\pm0.5 $  &\\
    \cline{2-6}
    &  $L_1$ &\multicolumn{2}{|l|}{$10.10\pm0.2$} & $\bm{ 69.86\pm1.0}$ & 95   \\
    &$L_2$ &\multicolumn{2}{|l|}{$ 8.86\pm0.2$}  & $17.10\pm4.3 $  &\\
    & No-regularization &\multicolumn{2}{|l|}{$\bm{8.29\pm0.2}$}  & $1.86\pm0.3 $  &\\
    \hline
\end{tabular}

\begin{tabular}{ c|c|l|l|l|l|l|c }
\multicolumn{6}{c}{}\\
\scriptsize{Model} & \scriptsize{Reg} &\multicolumn{2}{|l|}{\scriptsize{seg score (rand)$\uparrow$}} &\multicolumn{2}{|l|} {\scriptsize{segm score (MI)}}& \scriptsize{\% Pruned Nodes}& \scriptsize{\% Pruned Weights}\\
    \hline
    U-Net& $L_1$  
    &\multicolumn{2}{|l|}{ $94.48\pm 0.7$} &\multicolumn{2}{|l|}{$97.43\pm0.1$} 
    & $1.45\pm 0.2$ & $95$\\
    &$L_2$ 
    &\multicolumn{2}{|l|}{$\bm{ 95.81\pm0.7}$} 
    &\multicolumn{2}{|l|}{$\bm{98.12\pm0.2 }$}   &{$\bm{68.05\pm5.9}$} & \\
    & Elastic-net   
    &\multicolumn{2}{|l|}{$94.25\pm0.6$}  &\multicolumn{2}{|l|}{$97.39\pm0.7$} 
    & $3.46\pm2.5$ & \\
    & No-regularization  &\multicolumn{2}{|l|}{$94.28\pm0.8$}   &\multicolumn{2}{|l|}{$97.72\pm0.1$}  
    & $0$ & \\
    \hline

\scriptsize{Model$\uparrow$}& \scriptsize{Reg}&
\scriptsize{precision$\uparrow$}& \scriptsize{recall$\uparrow$}& \scriptsize{f1-score$\uparrow$}& \scriptsize{accuracy$\uparrow$}& 
\scriptsize{\% Pruned Nodes$\uparrow$ }& 
\scriptsize{\% Pruned Weights$\uparrow$}\\
\hline

    DarkCovidNet
    &$L_1$  
    & $88.67 \pm 0.4$
    & $82.33 \pm 1.4$ 
    & $84.50 \pm 0.7$ 
    & $84.00 \pm 0.1$
    & $68.87 \pm 0.6$ & 97\\
    
    &$L_2$  
    & $88.28 \pm 0.5$
    & $81.72\pm1.3$
    & $84.17 \pm 0.9$
    & $84.00	\pm0.7$
    & $40.07 \pm 8.4$
    & \\
    &No-regularization 
    & $88.89 \pm1.6$
    & $	83.00\pm1.4$ 
    & $85.11\pm1.1$ 
    & $84.33\pm2.0$
    & $0\pm0$ & \\
    \hline

\end{tabular}
\end{center}
\end{table*}

\begin{figure*}
\begin{tabular}{lll}
\hspace{-0.6cm}
\includegraphics[trim={1cm 0cm 1cm 0cm},clip,width=0.312\linewidth]{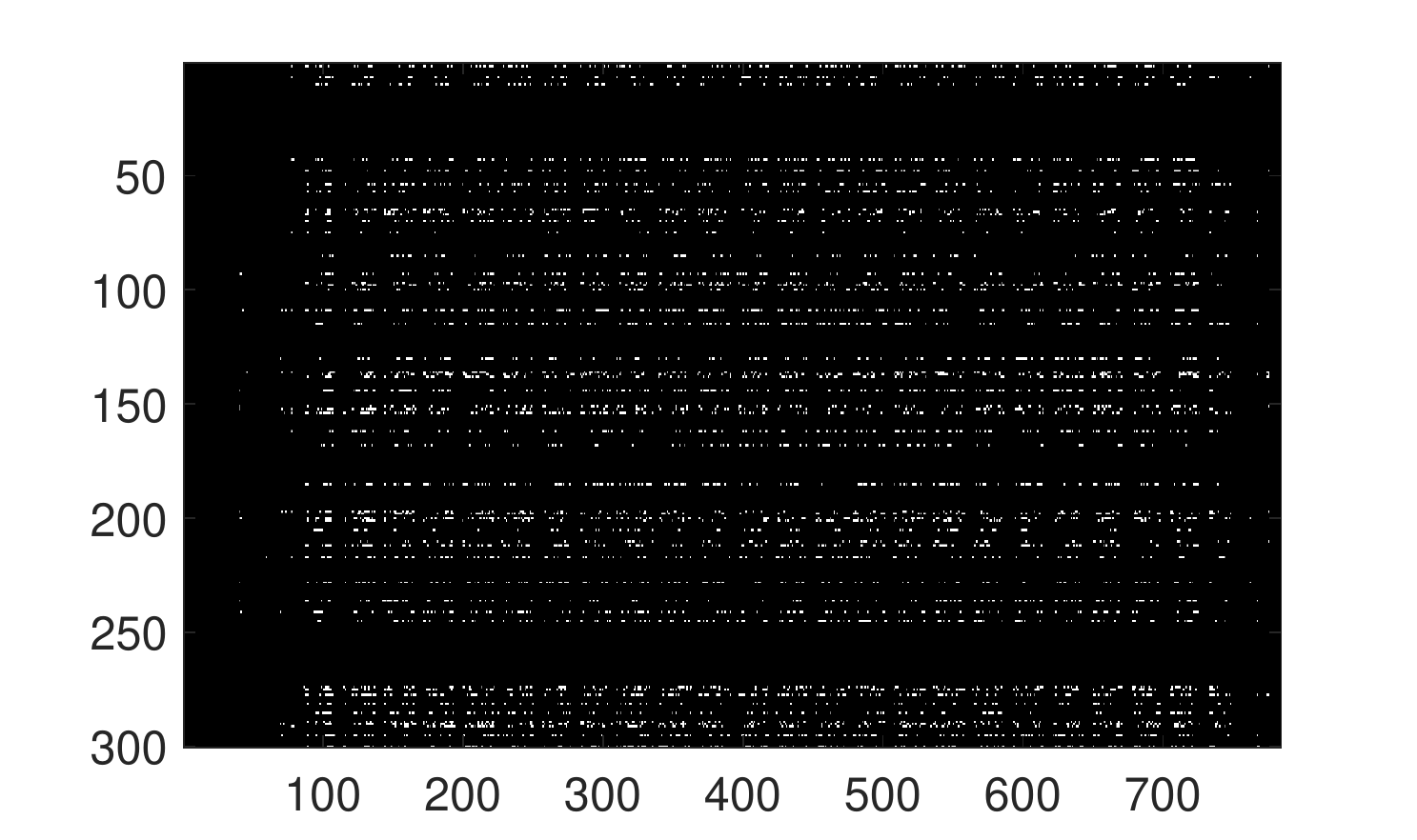}
\includegraphics[trim={1cm 0cm 1cm 0cm},clip,width=0.312\linewidth]{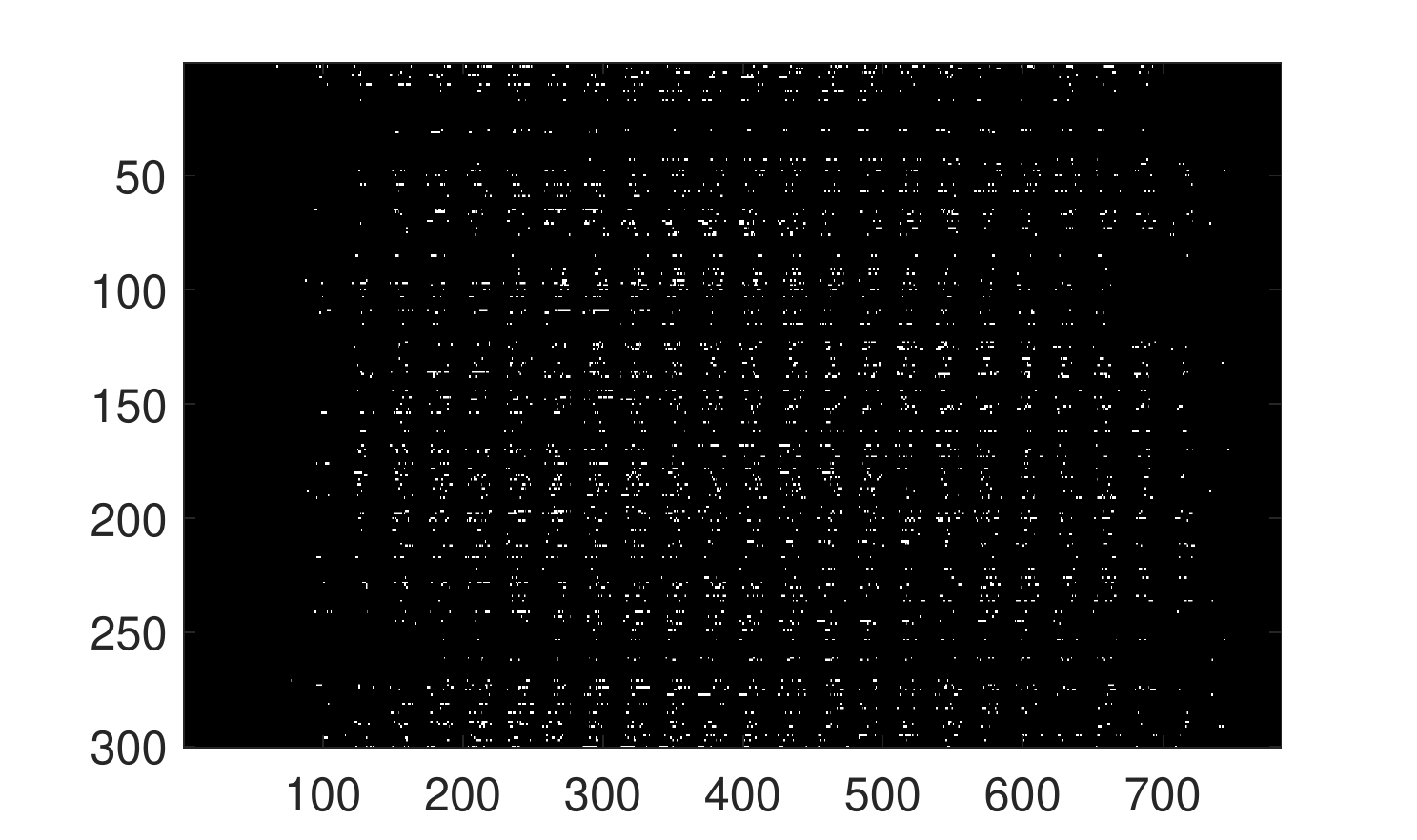}
\includegraphics[trim={1cm 0cm 1cm 0cm},clip,width=0.312\linewidth]{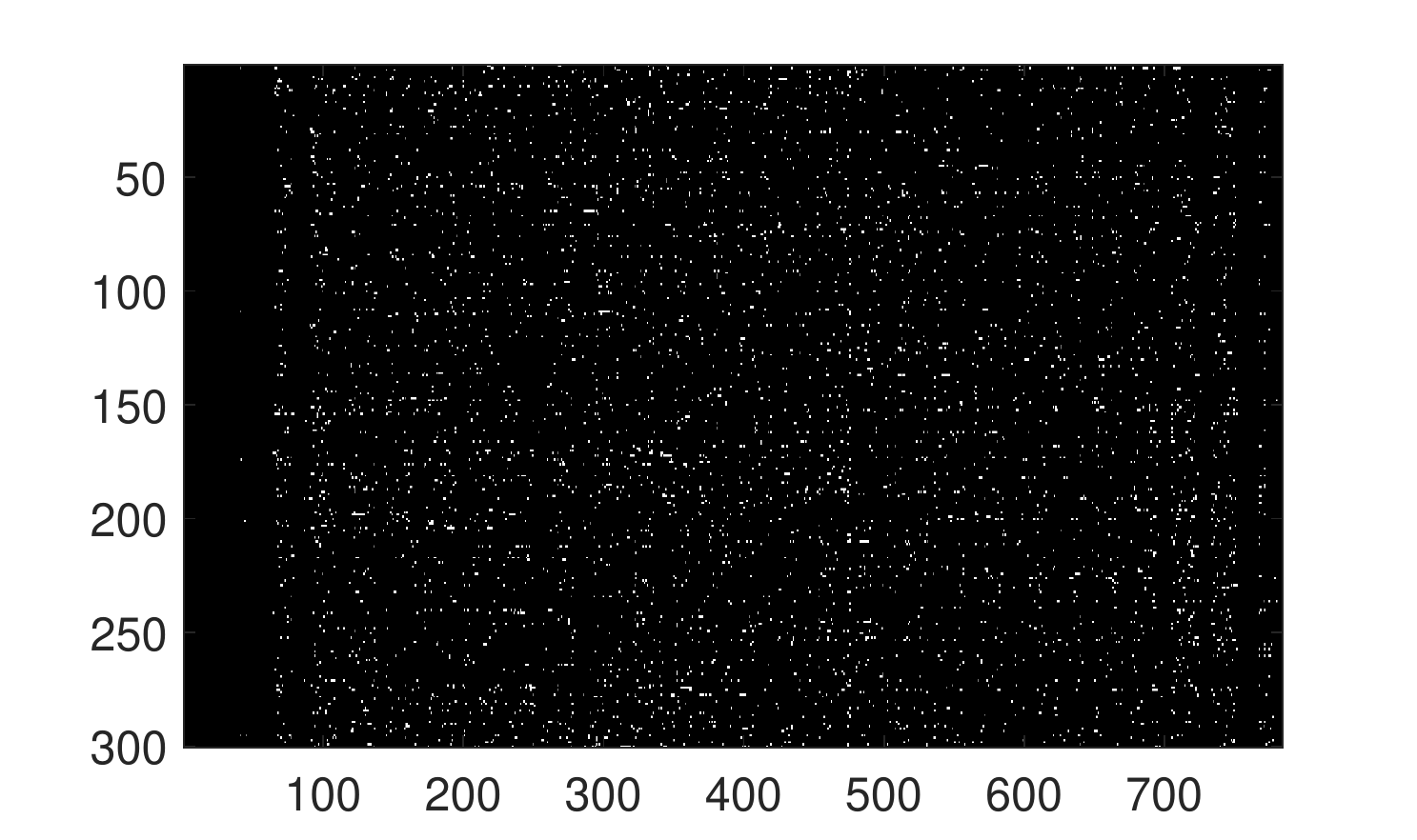} \\
\hspace{-0.6cm}
\includegraphics[trim={1cm 0cm 0.8cm 0cm},clip,width=0.31\linewidth]{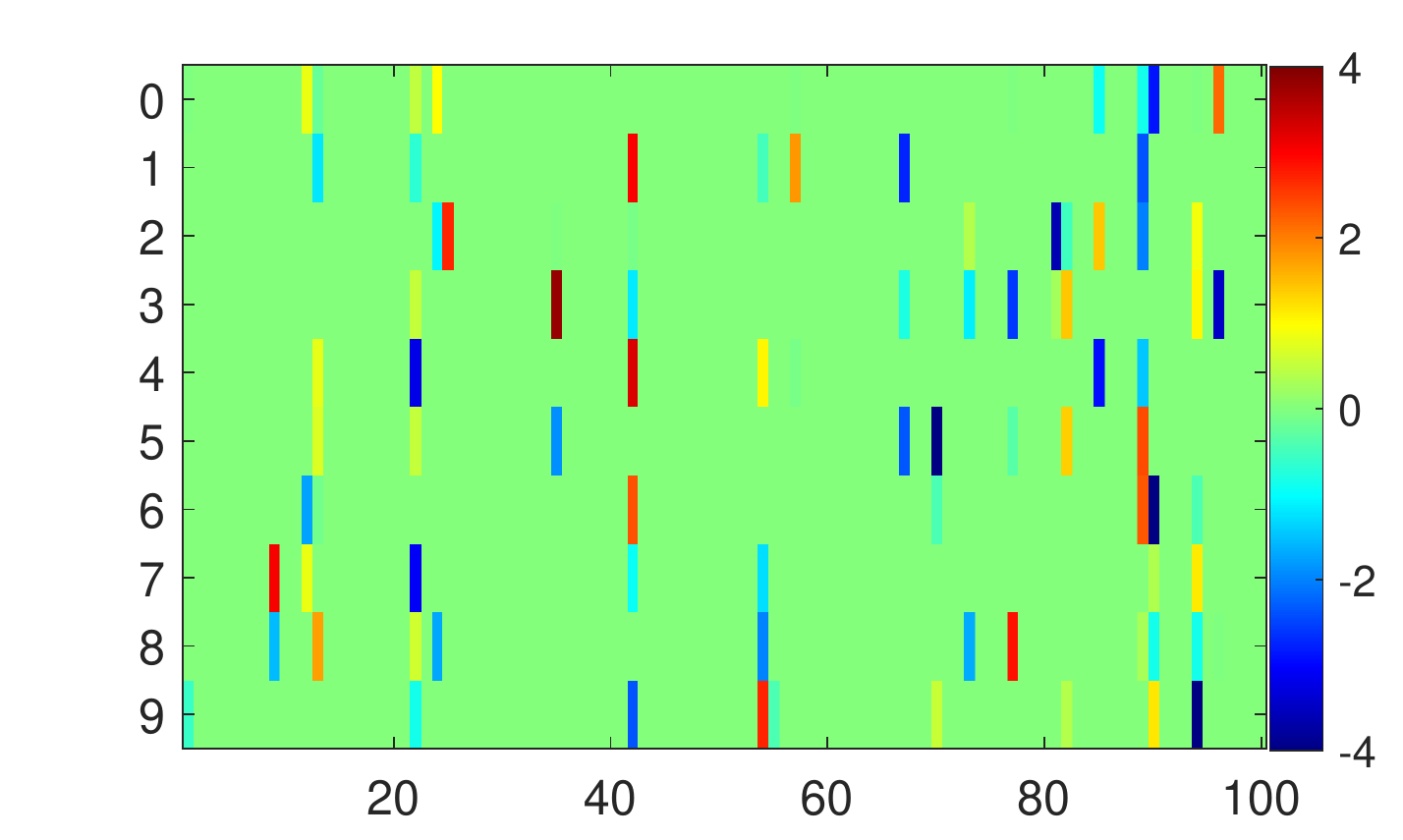}
\includegraphics[trim={1cm 0cm 0.8cm 0cm},clip,width=0.31\linewidth]{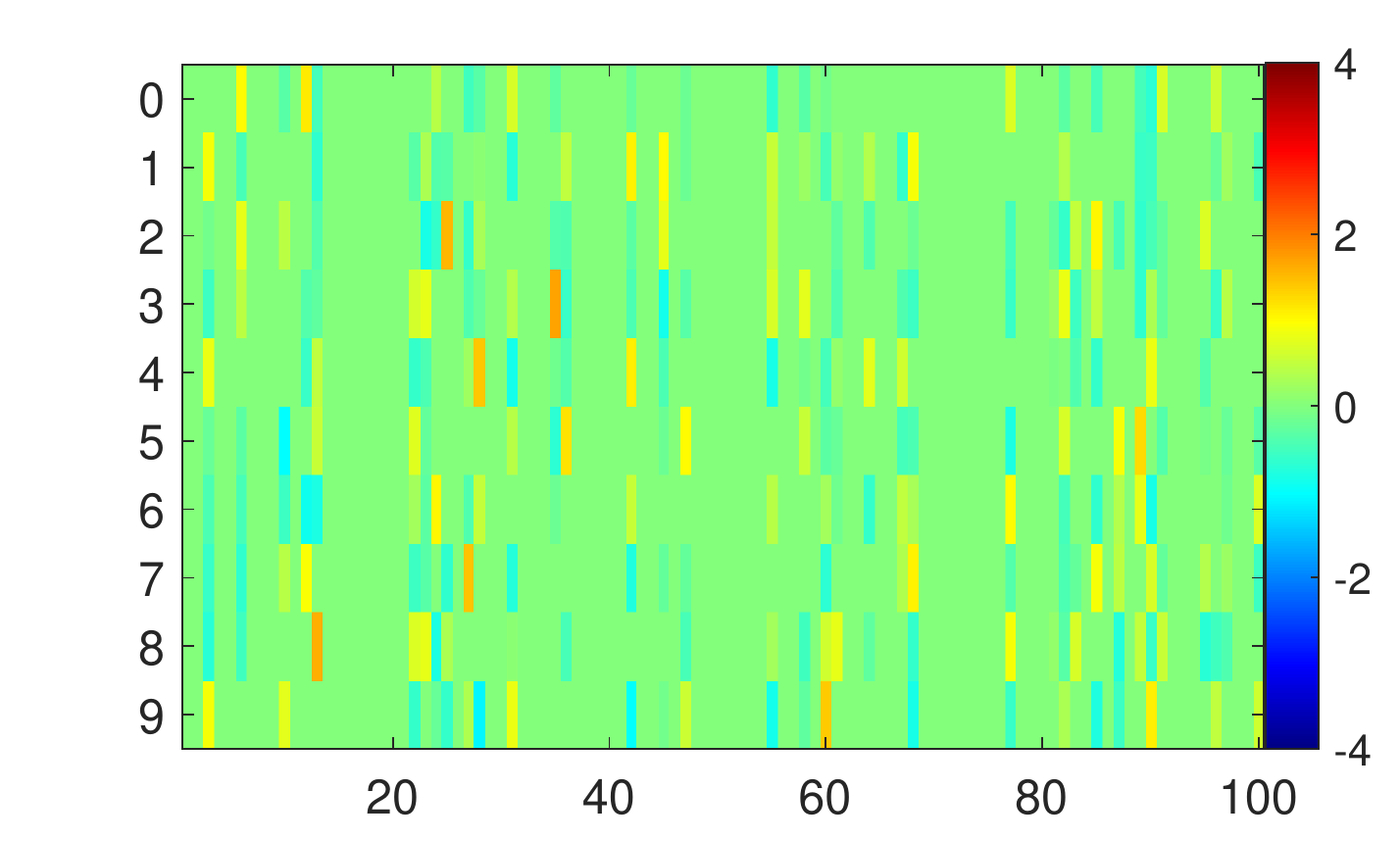}
\includegraphics[trim={1cm 0cm 0.8cm 0cm},clip,width=0.31\linewidth]{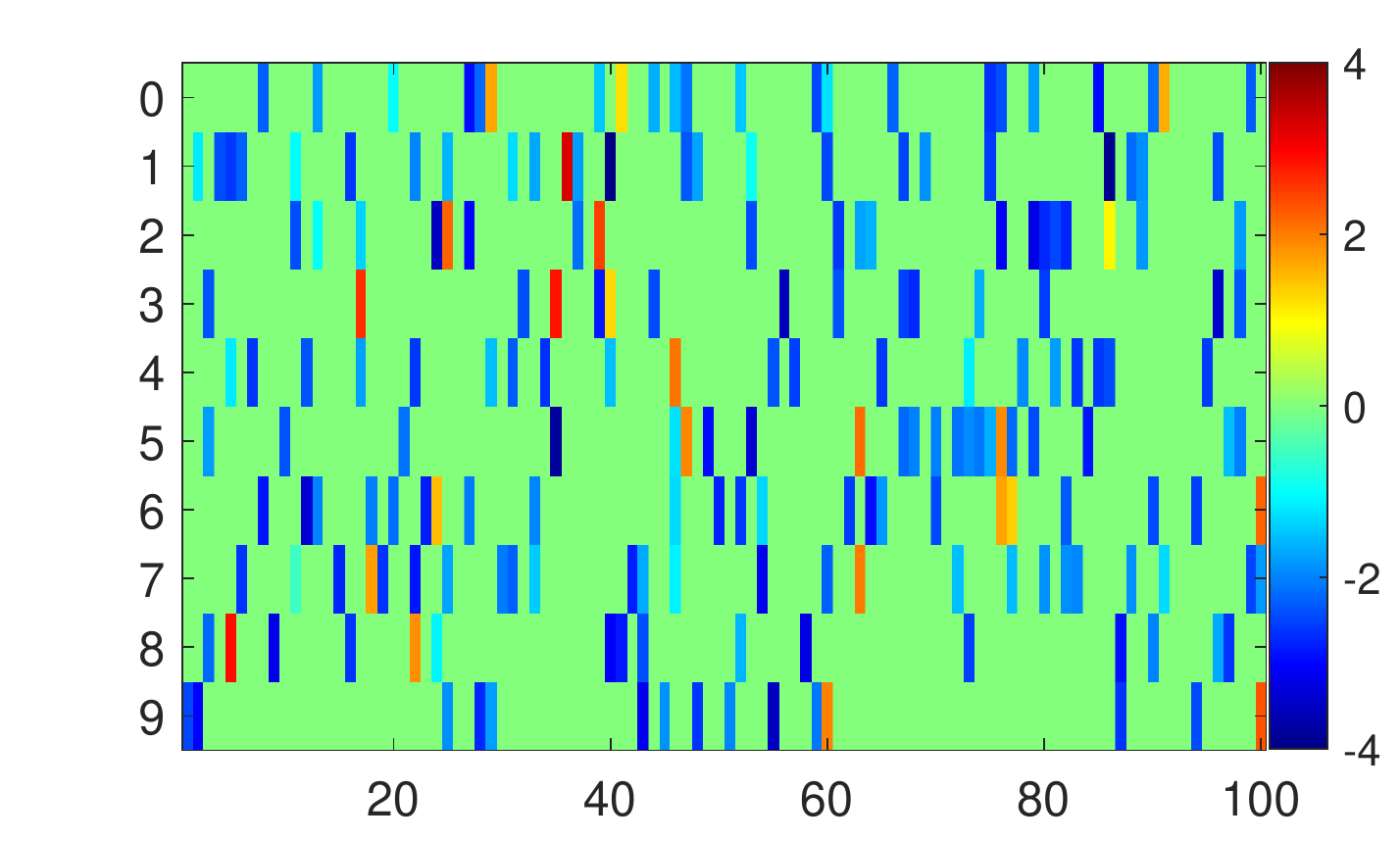} \\
\hspace{1.1cm}\hspace{0.9cm}$L1$\hspace{5cm}$L2$\hspace{3.8cm} w/o regularization 
\end{tabular}
\caption{Visualization of the remaining connections between consecutive layers in MLP-300-1000 after pruning with $L_1$ and $L_2$ (left, middle) and without (right) regularization. The first row in the figure presents binary maps of the remaining connections between the input layer (784 nodes) and the first hidden layer (300 nodes). The second row is a visual presentation of the weights (color coded) of the edges connecting the second hidden layer (100 nodes) to the output layer (10 nodes).}
\label{fig:connectionsMaps}
\end{figure*}

\begin{figure}[ht!]
\begin{center}
\centerline{\includegraphics[width=9.5cm]{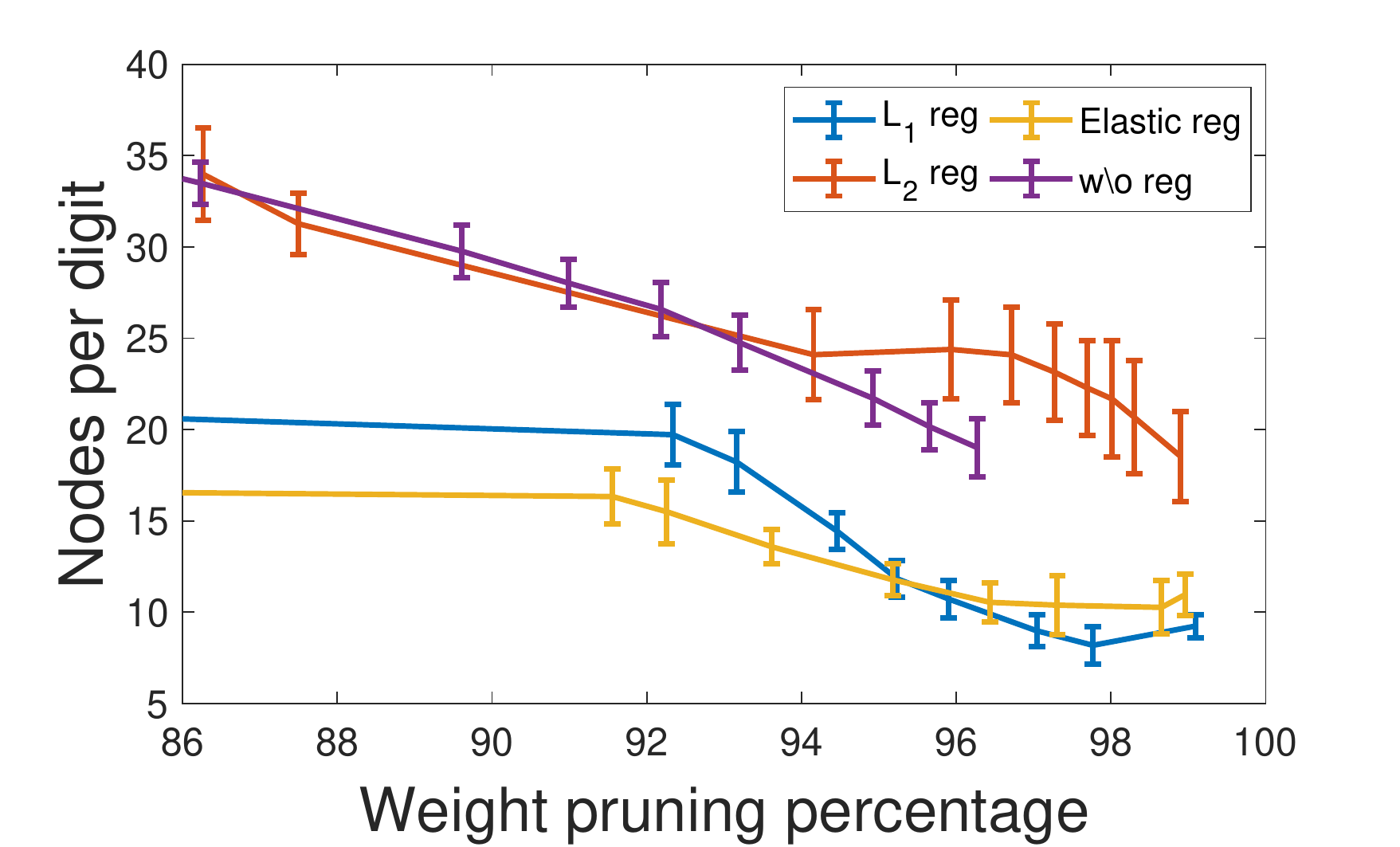}}
\caption{ The average number of nodes representing an MNIST digit as a function of the weight pruning percentage for MLP-300-100. Results are shown for pruning with $L_1$, $L_2$ and elastic-net as well as  without regularization.}
\label{fig:Digits}
\end{center}
\end{figure}

{\bf MLP-300-100 for MNIST classification:}
{The MLP-300-100 is composed of an input layer with 784 nodes, two fully connected hidden layers, termed FC1 and FC2 containing 300 and 100 nodes, respectively, and an output layer with 10 nodes, where each node represents a digit.   
Figure~\ref{fig:connectionsMaps} presents a visualization of the weights of the edges connecting between the consecutive MLP layers. The first row in the figure presents binary maps of the remaining connections between the input layer (784 nodes) and the first hidden layer (300 nodes) after pruning with either the $L_1$(left) or $L_2$ (middle) regularization terms or without any regularization (right). While the number of removed connections (black) seems to be similar in all the plots, the patterns of the remaining connections for the regularized pruning appear to be much more coherent, which means that entire rows and columns (nodes) were removed. On the other hand, when no regularization was applied, the pattern of remaining connections was random. The second row presents the weights (color coded) of the edges connecting the second hidden layer (100 nodes) to the output layer (10 nodes) visually. Note that for the regularized pruning, very few nodes were required to represent each of the output digits. In addition, as expected, the magnitudes of the remaining connections for the pruning with $L_2$ regularization were much lower. The weights' color map for unregularized pruning was much denser, with many more active nodes (columns) in general. Note however (see also first row in table ~\ref{tbl:Digits} and figure ~\ref{fig:Digits}) that the average number of nodes require to represent each digit is similar to one obtained using pruning with $L_2$ regularization. 
The maps presented in the figure correspond to the results shown in Table~\ref{tbl:MLPResults} when pruning 97\% of the weights.} 

{In the next experiment we examined the average number of nodes required to represent an MNIST digit.
Note that in the original unpruned network, most of the 100 nodes in FC2 were connected to all 10 nodes in the output layer, where each node represents a single digit. 
Figure~\ref{fig:Digits} and  Table~\ref{tbl:Digits} present the average number of nodes in FC2 that remained connected to each of the nodes in the output layer for different regularization schemes. These average numbers are presented for the case of 95\% pruned weights (Table~\ref{tbl:Digits}) and as a function of the percentage of pruned weights (Figure~\ref{fig:Digits}). The last row in Table \ref{tbl:Digits} presents the percentages of pruned nodes in FC2 when 95\% of the pruned weights were removed. Note that we ran each experiment 4 to 6 times so that the tables include standard deviation values, shown as error-bars in the figures.} 

\begin{table}[ht!]
\begin{center}
\caption{The average number of nodes in FC2 that remained connected to a node in the output layer (nodes per digits) and the percentage of pruned nodes in FC2 for MLP-300-100 with different regularization schemes. $\downarrow$ lower is sparser, $\uparrow$ higher is sparser.}
\label{tbl:Digits}
\small
\centering
\begin{tabular}{ l|l|l }
 \hline
 \scriptsize{Regularization}
& \scriptsize{\# nodes per digit}
& \scriptsize{\% pruned nodes in FC2}
\\
\hline
$L_1$& $11.8\pm1.0$ &  $84.7\pm1.4$\\
\hline
$L_2$& $24.4\pm2.7$ &  $64.8\pm3.1$\\
\hline
elastic-net& $11.8\pm0.9$ &  $80.4\pm1.9$\\
\hline
w/o reg.& $21.7\pm1.5$  &  $3.5\pm2.1$\\
\hline
\end{tabular}
\end{center}
\end{table}

\begin{figure*}
\begin{tabular}{c}
\includegraphics[width=0.9\linewidth]{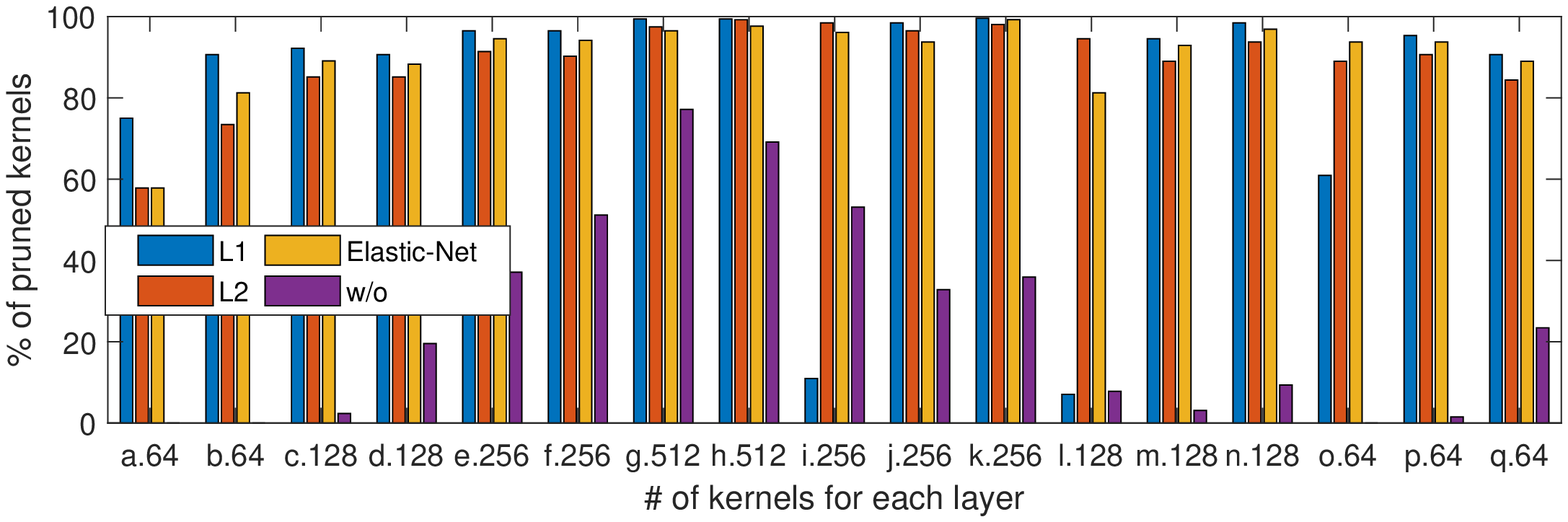}\\
\hspace*{2cm}
\includegraphics[width=0.8\linewidth]{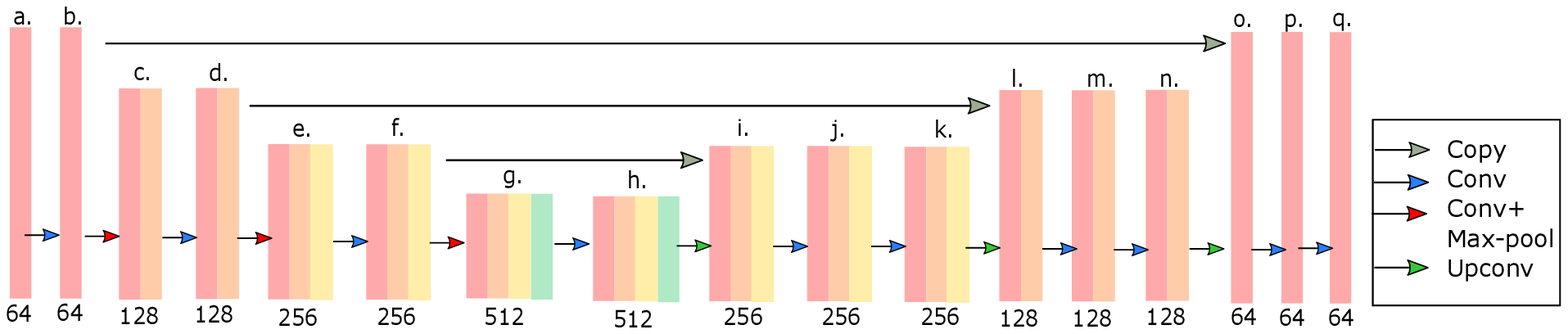}
\end{tabular}
\caption{{\bf The U-Net architecture along with its respective bar plot presenting the percentage of pruned kernels in each of its layers.}
Lower panel: visualization of the U-Net architecture. 
The U-Net is composed of convolutional encoder layers of decreasing size followed by convolutional decoder layers of increasing size. In addition, corresponding encoder-decoder layers are connected with skip connections. The number of kernels in each layer is indicated underneath the bar that represents it. \\
Upper panel: 
percentage of pruned kernels in each layer for the U-Net architecture for $L_1$, $L_2$, elastic-net regularization and no regularization. (top image).\\
Modifies U-net architecture were each box corresponds to a multi-channel feature map, with the number of channels indicated below the box. Arrows in different colors denote different operations. The gray arrows correspond to copied feature maps. (bottom image).}
\label{fig:UnetLayers}
\end{figure*}

{\bf U-Net for cell segmentation in microscopy images:}
{Figure~\ref{fig:UnetLayers} presents the U-Net architecture (lower panel) along with the respective bar plot (upper panel) listing the percentages of pruned kernels in each layer. Recall that the U-Net was diluted by direct pruning of its weights. In the presence of $L_1$ (blue), $L_2$ (red) and elastic-net (yellow) regularizations throughout the weight pruning process, complete U-Net's kernels were pruned as well. As expected, the node pruning percentages were much lower when no regularization was applied (purple bars).      
Table ~\ref{tbl:UnetLayers} presents the respective percentages of pruned weights and kernels in the entire network along with the segmentation scores for the test data.}  

\begin{table}[ht]
\begin{center}
\caption{Percentages of the pruned kernels and weights in the entire network along with the segmentation (seg.) scores in the presence of $L_1$, $L_2$, or elastic-net regularization terms or without any regularization.}
\label{tbl:UnetLayers}
\small
\centering
\begin{tabular}{ l|l|l|l|l }
 \hline
\scriptsize{}
& \scriptsize{$L_1$}
& \scriptsize{$L_2$}
& \scriptsize{Elastic-net }
& \scriptsize{w/o reg }
\\
\hline
\% pruned kernels & $75.4$ & $82.0$ & $82.3$ & $ 36.5$\\
\hline
\% pruned weights & $99.9$ & $99.7$ & $99.8$ & $99.8$\\
\hline
seg. score (MI) & $97.6$ & $97.7$ & $97.3$ & $ 97.5$\\
\hline
seg. score (Rand) & $94.9$ & $95.1$ & $93.9$ & $95.0$\\
\hline
\end{tabular}
\end{center}
\end{table}

{\bf DarkCovidNet Model for automatic COVID-19 detection using raw chest X-ray images:}

\begin{figure*}
\begin{tabular}{c}
\includegraphics[width=1\linewidth]{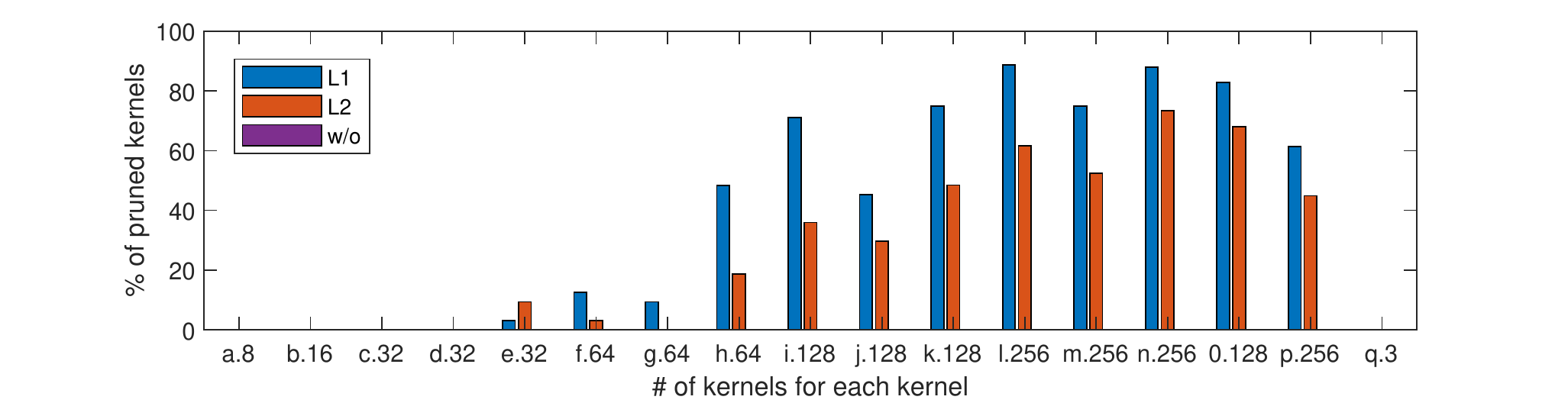}\\
\includegraphics[width=0.8\linewidth]{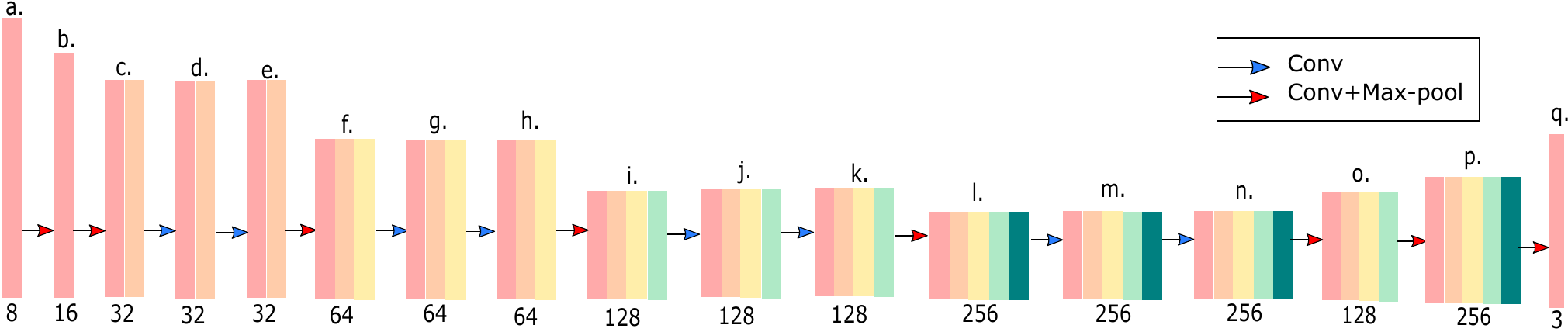}
\end{tabular}
\caption{{\bf The DarkCovidNet architecture along with its respective bar plot presenting the percentage of pruned kernels in each of its layers.}
Lower panel: visualization of the DarkCovidNet architecture. 
TheDarkCovidNet is composed of convolutional layers of decreasing size. The number of kernels in each layer is indicated underneath the bar that represents it. \\
Upper panel: 
percentage of pruned kernels in each layer for the DarkCovidNet architecture for $L_1$, $L_2$ and no regularization. (top image).\\
DarkCovidNet architecture were each box corresponds to a multi-channel feature map, with the number of channels indicated below the box. Arrows in different colors denote different operations. (bottom image).}
\label{fig:CovidLayers}
\end{figure*}

{Figure~\ref{fig:CovidLayers} presents the DarkCovidNet architecture (lower panel) along with the respective bar plot (upper panel) listing the percentages of pruned kernels in each layer for weights pruning percentage of 97\%. In the presence of $L_1$ (blue), $L_2$ (red) regularizations throughout the weight pruning process, complete kernels of the deeper layers were pruned. This indicates that there is redundancy, especially in the deep layers. 
When no regularization was applied, the node pruning percentage is zero.}
\begin{figure}[t]
\centering
\begin{tabular}{ll}
\includegraphics[width=0.47\linewidth]{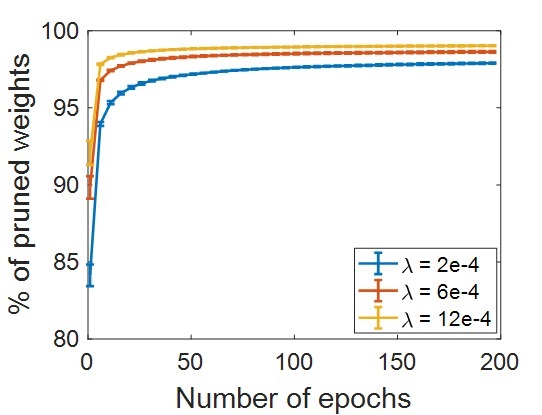} &
\hskip -4ex
\includegraphics[width=0.47\linewidth]{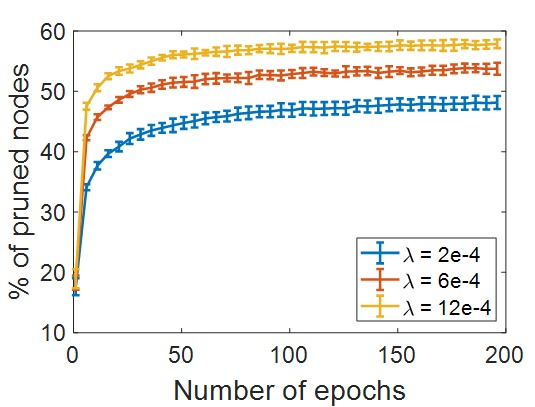} 
\end{tabular}
\caption{{\bf Pruning dynamics for MLP-300-100 applied to the MNIST classification.} Weight (left) and node (right) pruning percentages as a function of the number of epochs. Three sets of experiments were conducted using $L_2$ regularization with different values $\lambda = 2e\!-\!4,6e-4,12e-4$ in \textcolor{blue}{blue},~\textcolor{red}{red} and  \textcolor{yellow}{yellow}, respectively.}
\label{fig:lambda}
\end{figure}

\begin{table}
\begin{center}
\caption{Node and weight pruning percentages in MLP-300-100 when $L_2$ regularization was applied for different values of $\lambda$. The results refer to Figure~\ref{fig:lambda} in the main paper. }
\label{tbl:MLPlambda}
\small
\centering
\begin{tabular}{ c|c|c|c }
 \hline
 \scriptsize{}
& \scriptsize{$2e-4$}
& \scriptsize{$6e-4$}
& \scriptsize{$12e-4$ }
\\
\hline
\% pruned weights & $97.7\pm0.1 $ & $98.5\pm0.1$ & $98.9\pm0.1$  \\
\% pruned nodes & $48.2\pm1.1$ & $ 53.9\pm0.7$ & $57.7\pm0.9$ \\
\hline
\end{tabular}
\end{center}
\end{table}

\begin{figure*}
	\centering
	\begin{tabular}{ccccc}
	    & \hskip -2ex MLP-300-100 & \hskip -2ex VGG-like & \hskip -3ex U-Net & \hskip -2ex DarkCovidNet\\
        \begin{sideways}~~~pruned weights \end{sideways}&
        \hskip -2ex
        \includegraphics[width=0.25\linewidth]{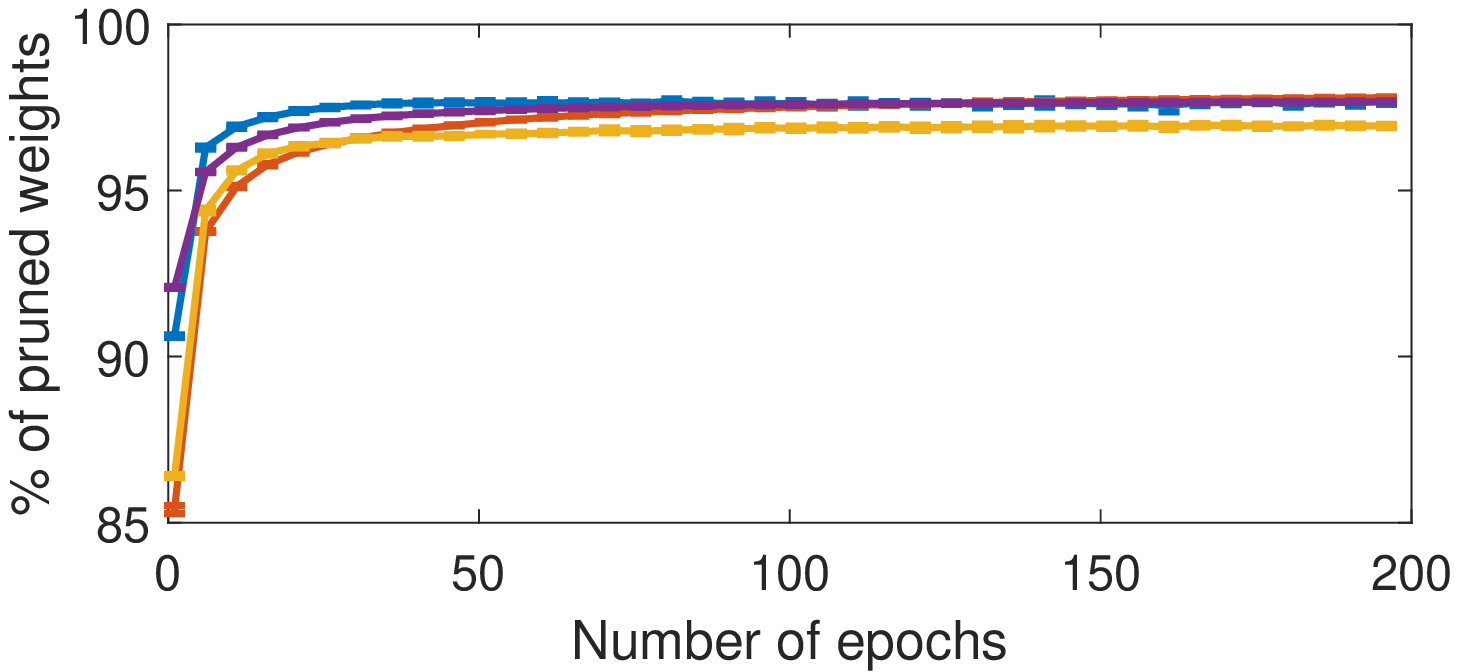} &
        \hskip -4ex
        \includegraphics[width=0.25\linewidth]{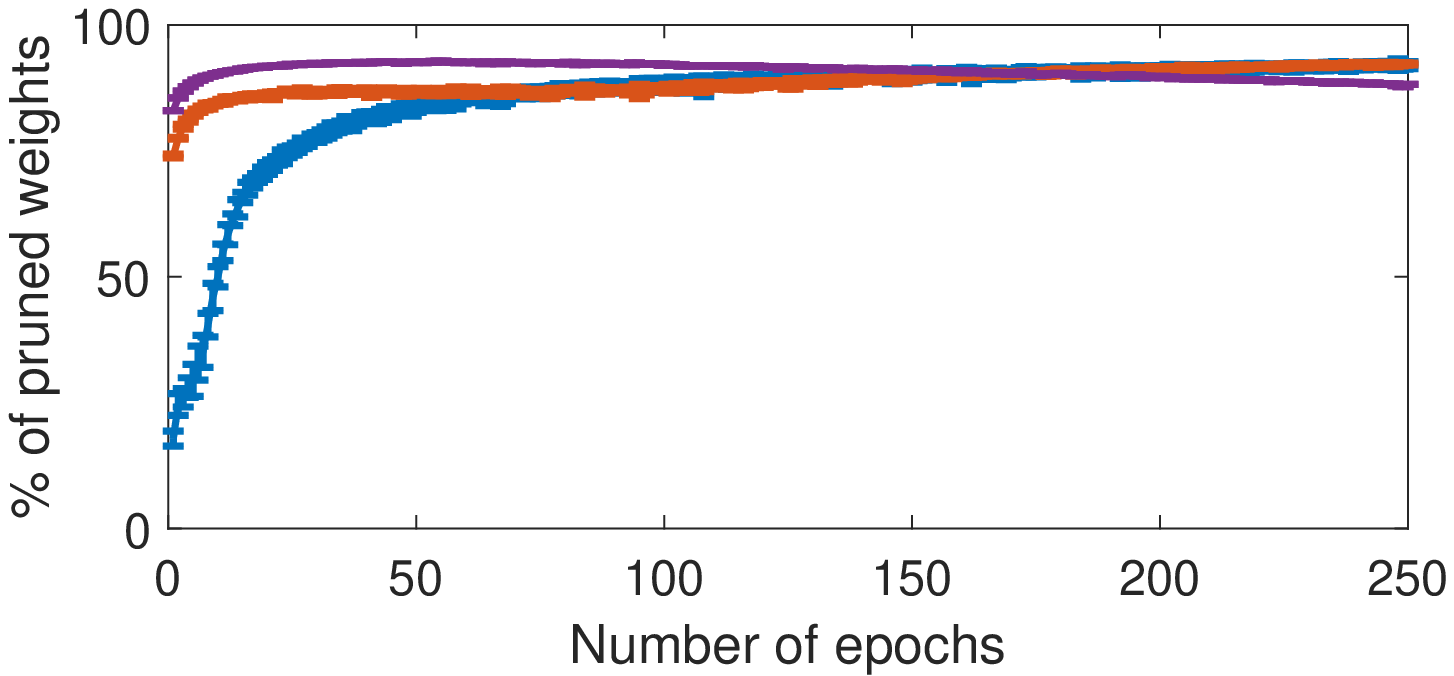}&
        \hskip -4ex
         \includegraphics[width=0.25\linewidth]{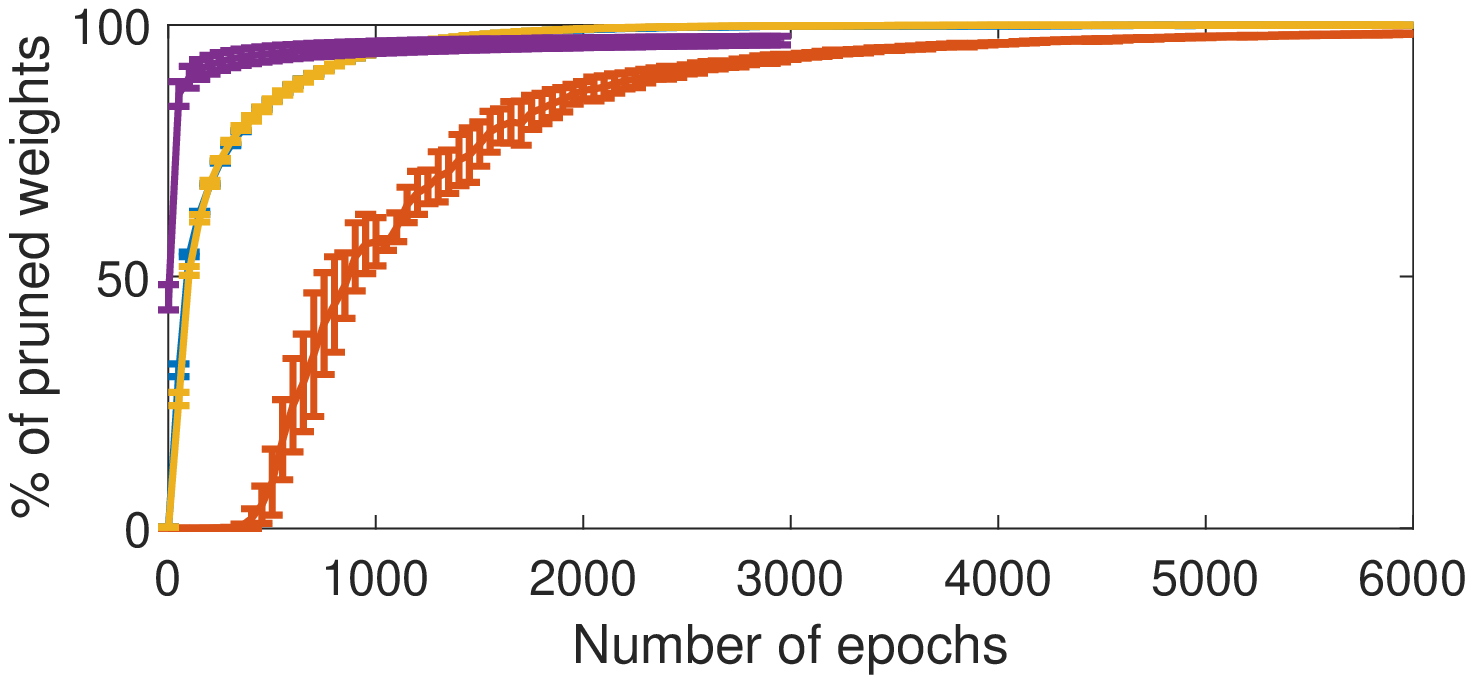}&
        \hskip -4ex
        \includegraphics[width=0.25\linewidth]{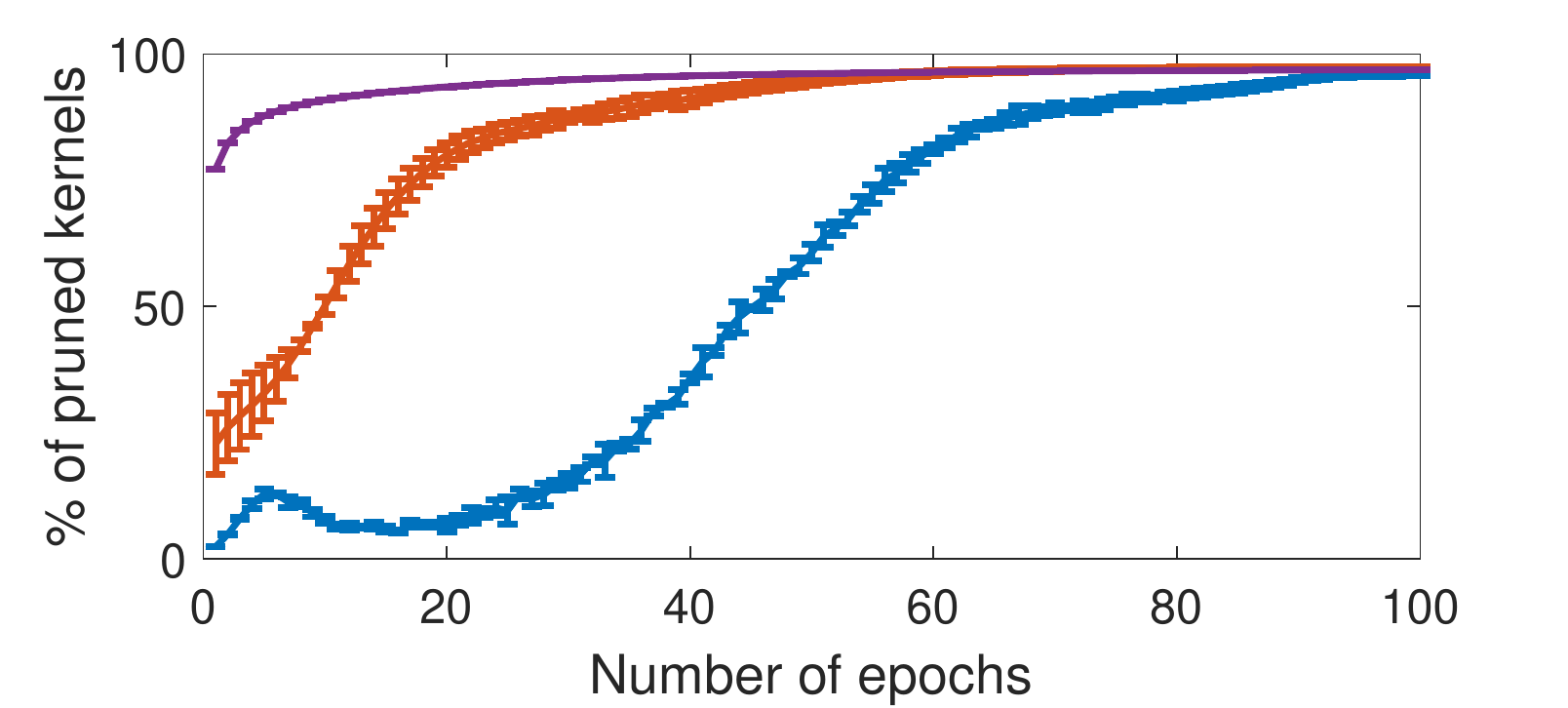}
        \\
        \begin{sideways}pruned nodes \end{sideways}&
        \hskip -2ex
        \includegraphics[width=0.25\linewidth]{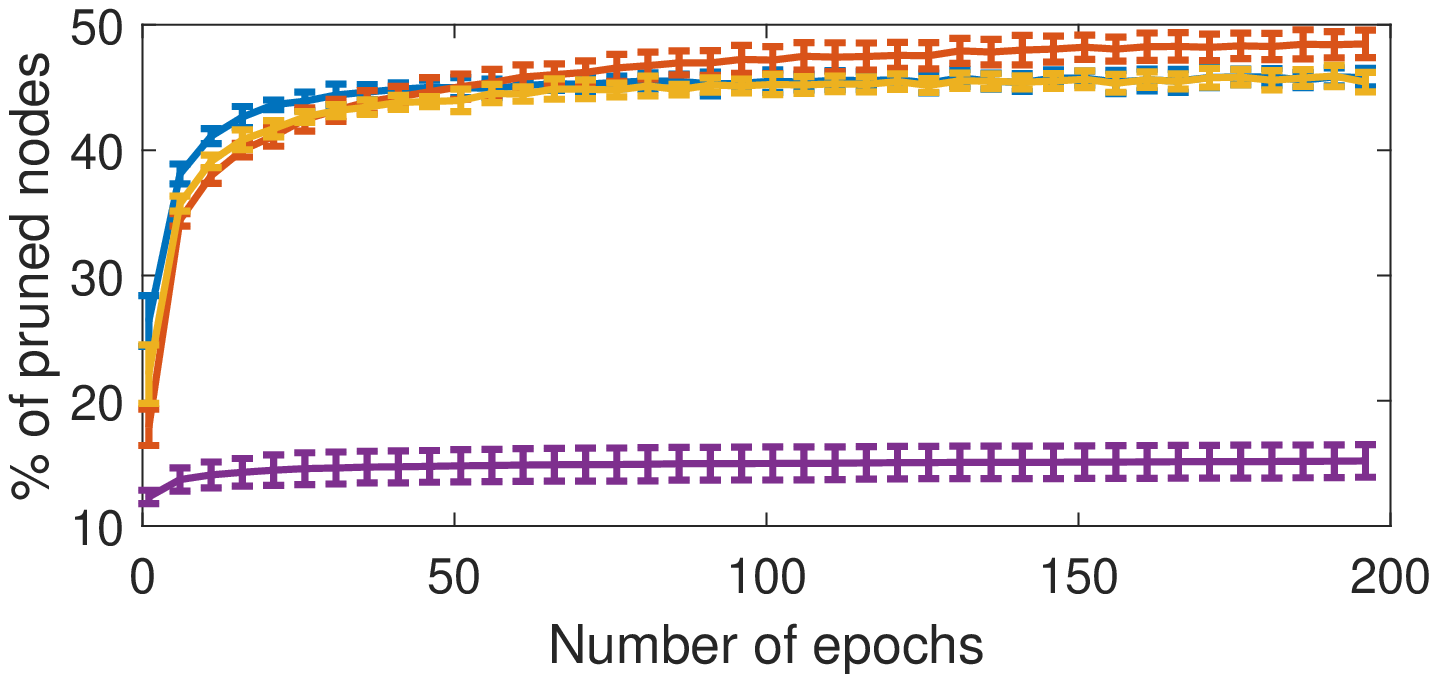}&
          \hskip -4ex
        \includegraphics[width=0.25\linewidth]{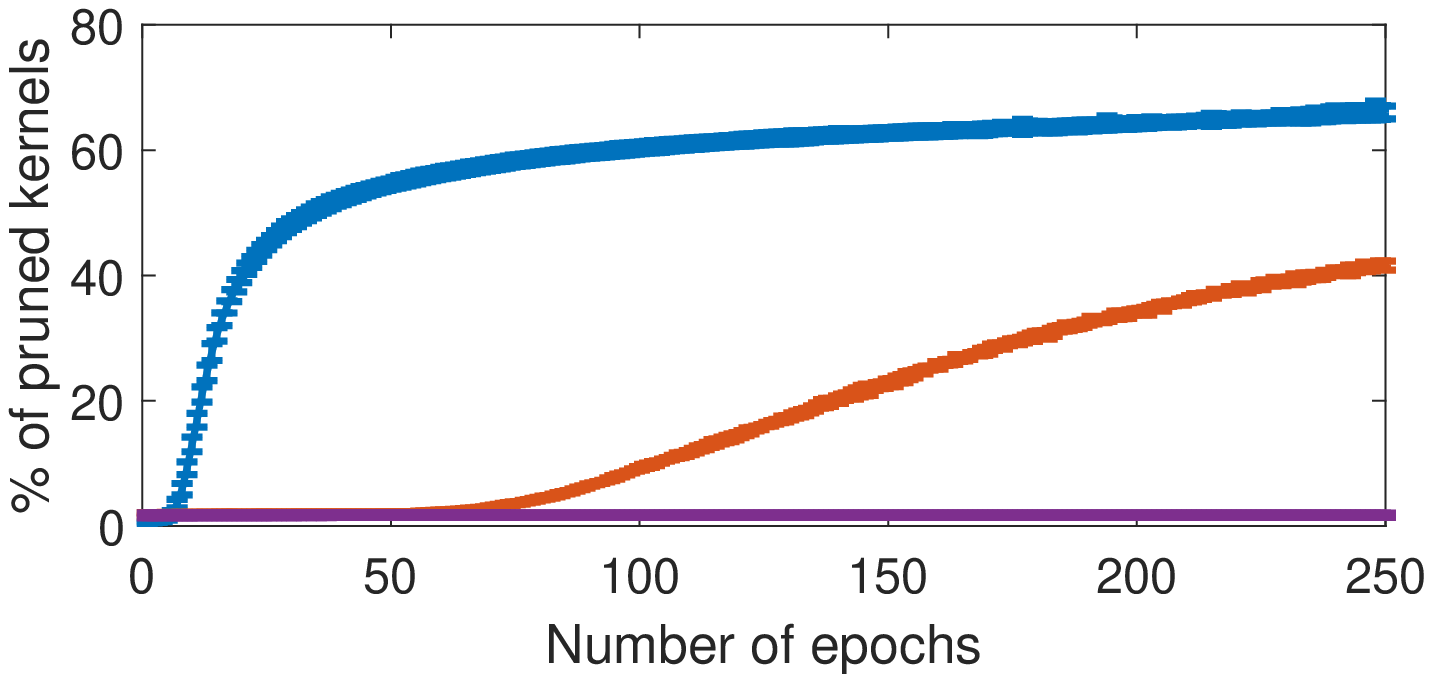}&
        \hskip -4ex
         \includegraphics[width=0.25\linewidth]{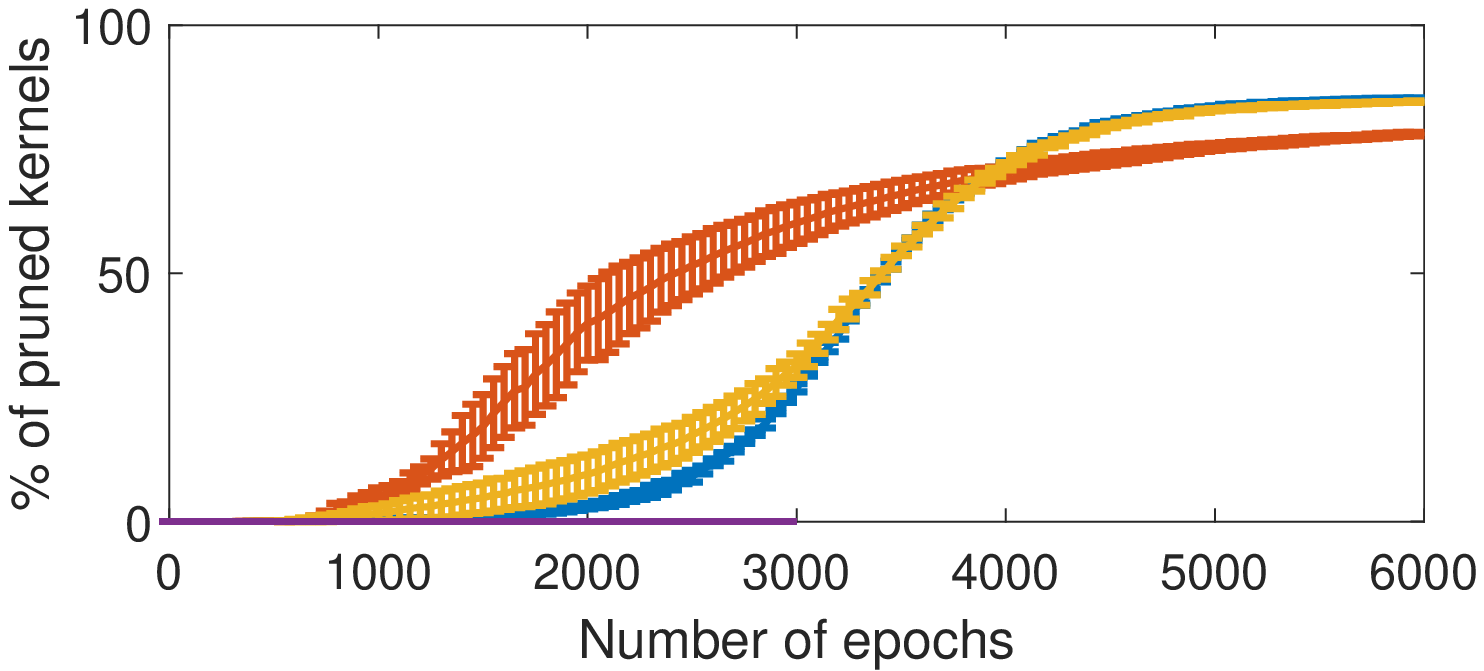}&
        \hskip -4ex
        \includegraphics[width=0.25\linewidth]{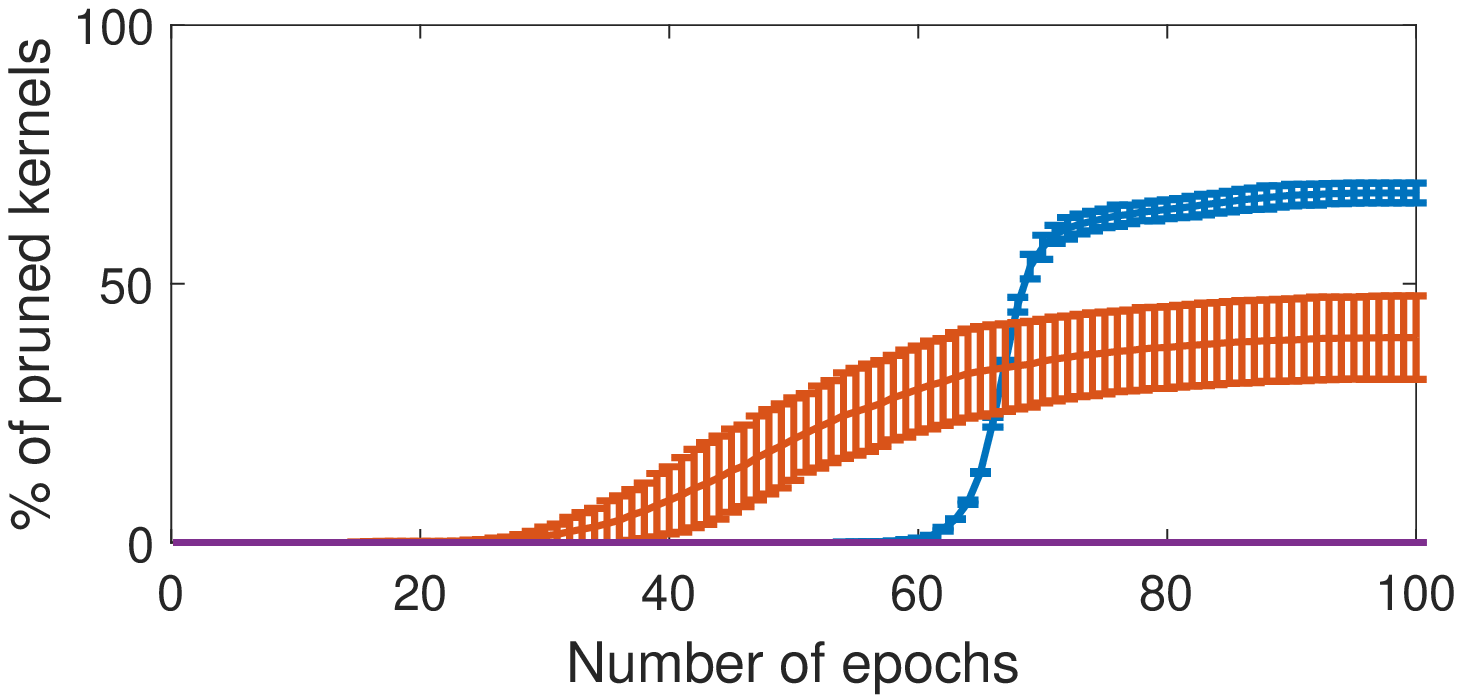}
        \\
        \end{tabular}
	\caption{Percentages of pruned weights (left) and pruned kernels/nodes (right) for MLP-300-100 (first row), for VGG-like (second row) , the U-Net (third row) and DarkCovidNet (forth row) calculated for each epoch. The plots are presented for different regularization terms: $L_1$ (\textcolor{blue}{blue}), $L_2$ (\textcolor{red}{red}) and  elastic-net (\textcolor{yellow}{yellow}) and with no regularization (\textcolor{violet}{purple}).}
	\label{fig:reg-terms}
\end{figure*}

\begin{figure*}
    \centering
        \begin{tabular}{ccccc}
        & MLP-300-100 &  VGG-Like & U-Net& DarkCovidNet \\
        \begin{sideways}~~~~~~~~~ No regularization \end{sideways}&
        \hskip -2ex
        \resizebox{\width}{3cm}{\includegraphics[width=0.25\linewidth]{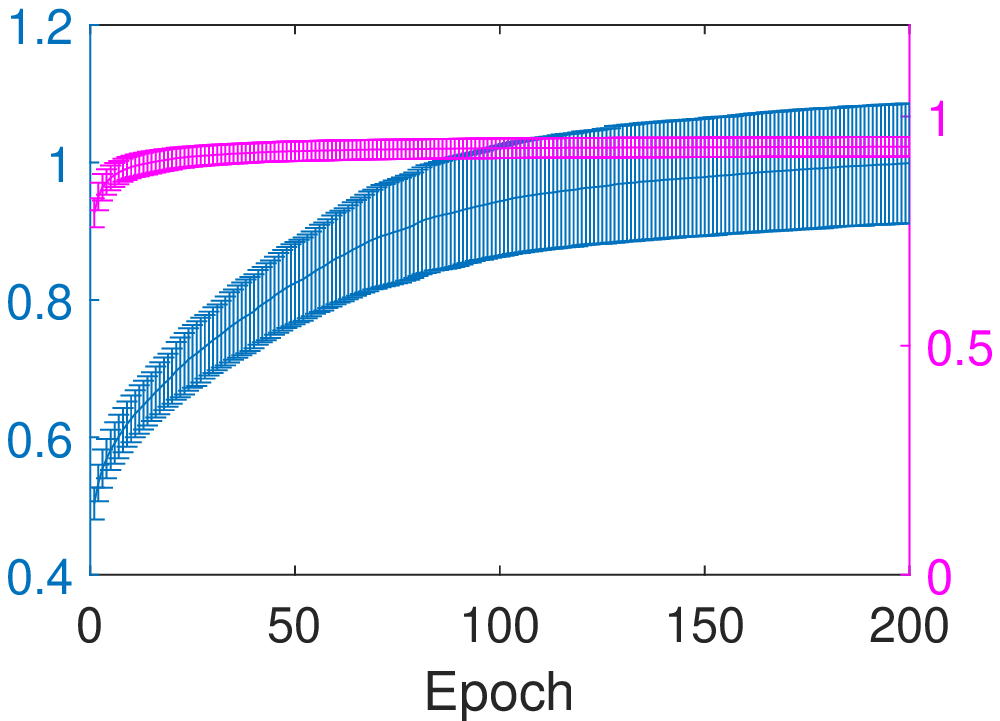}}&
        \hskip -4ex
        \resizebox{\width}{3cm}{\includegraphics[width=0.25\linewidth]{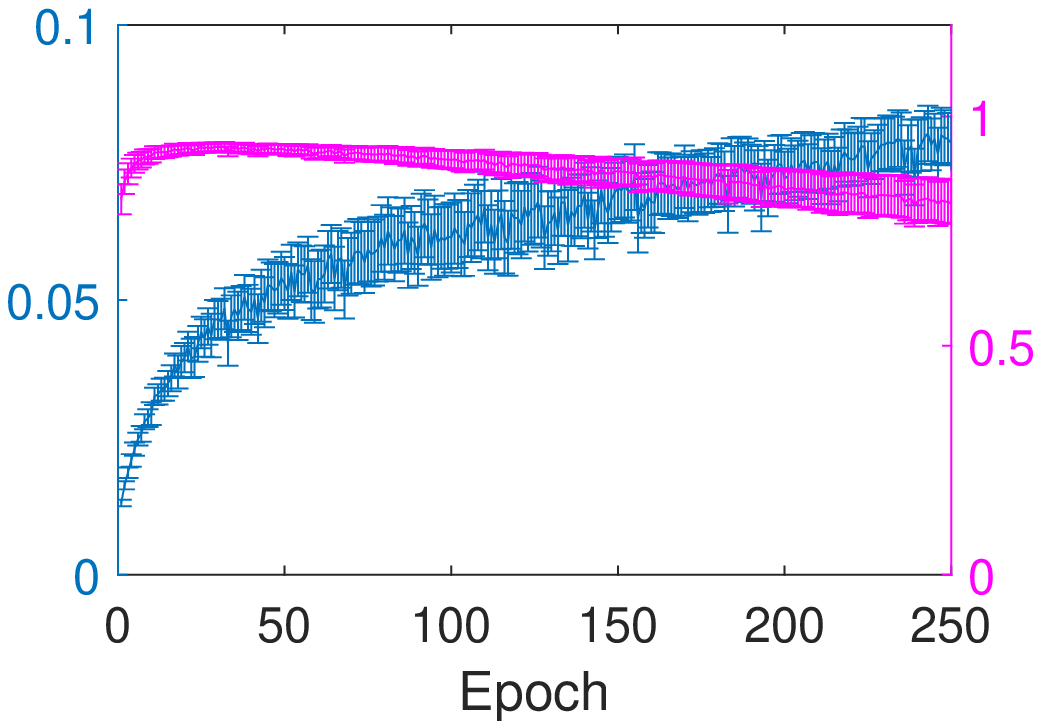}}&
        \hskip -4ex
        \resizebox{\width}{3cm}{\includegraphics[width=0.25\linewidth]{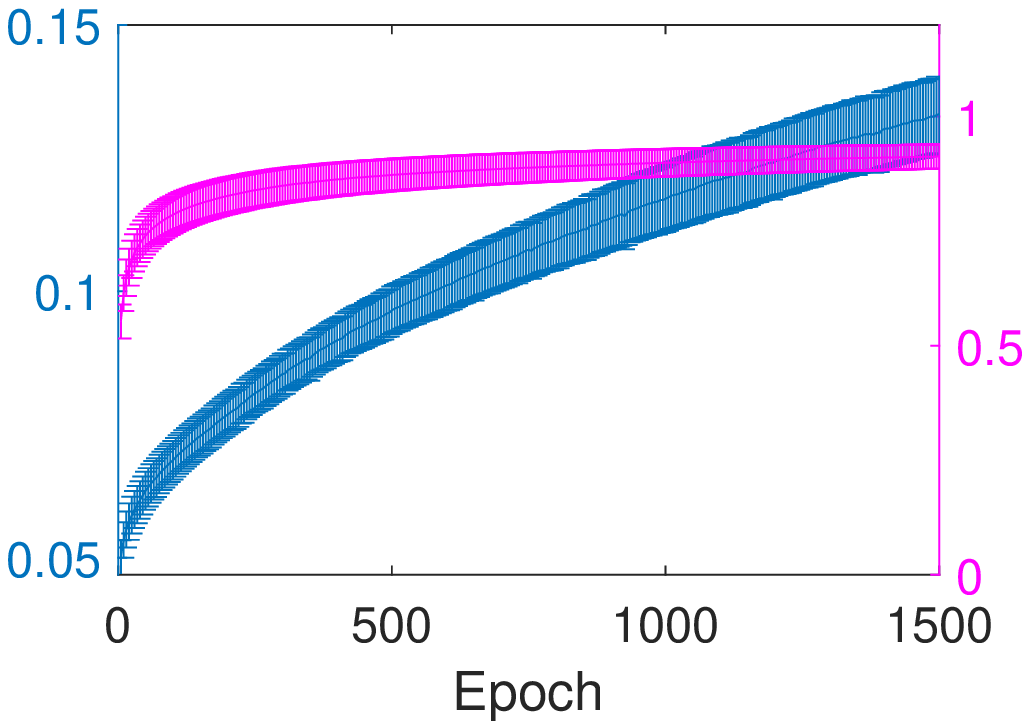}}&
        \hskip -4ex
        \resizebox{\width}{3cm}{\includegraphics[width=0.25\linewidth]{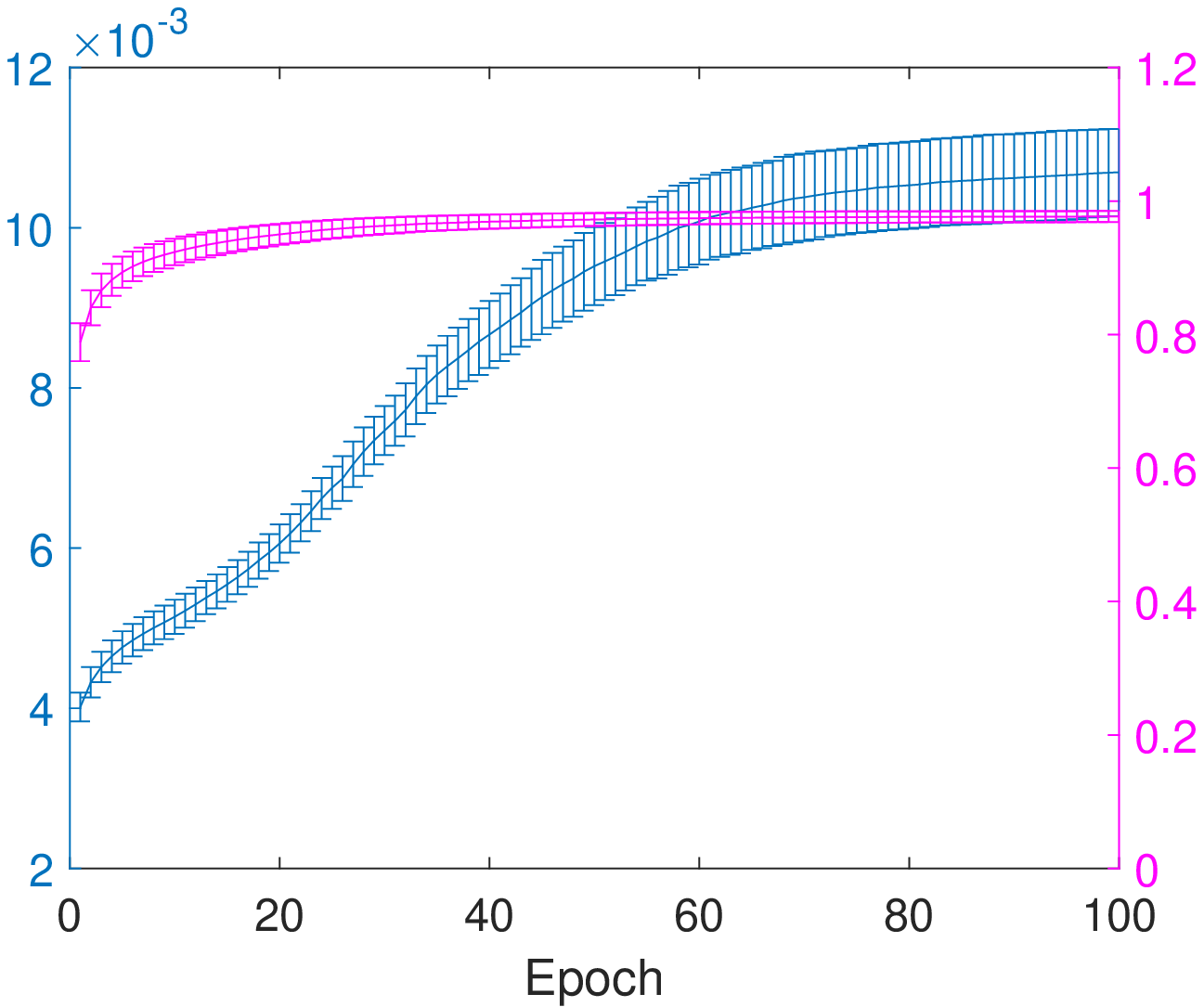}}\\ 
        \begin{sideways}~~~ $L_2,$  non-zeroed nodes \end{sideways}&
        \hskip -2ex
        \resizebox{\width}{3cm}{\includegraphics[width=0.25\linewidth]{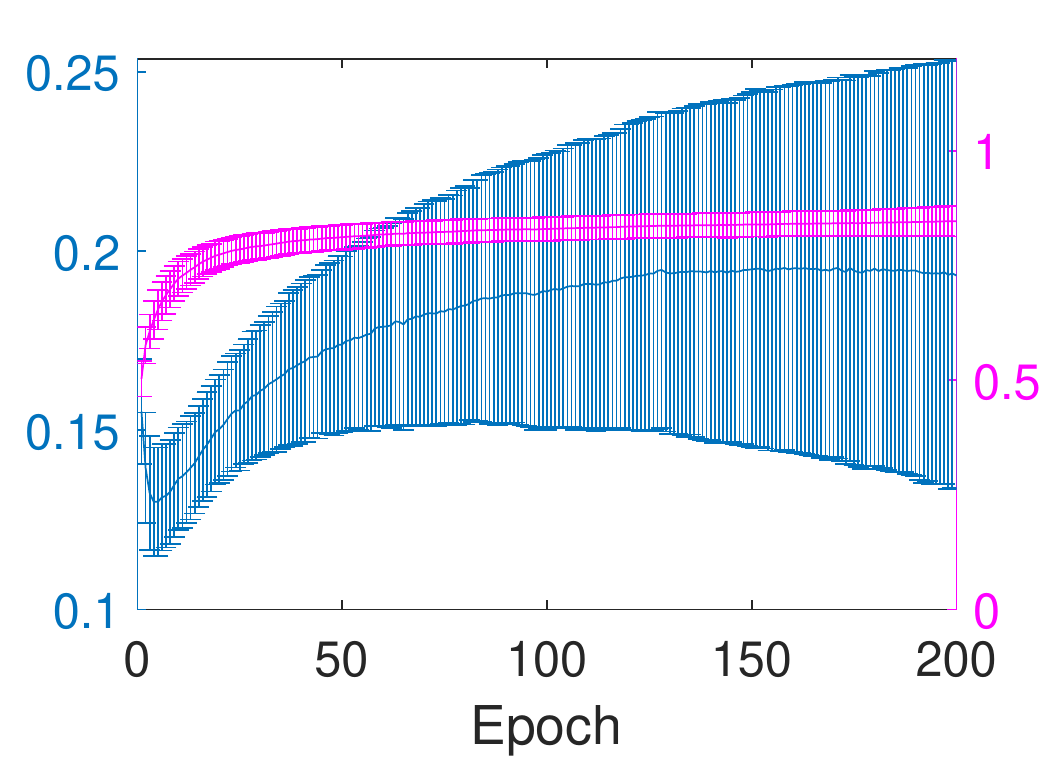}}&
        \hskip -4ex
        \resizebox{\width}{3cm}{\includegraphics[width=0.25\linewidth]{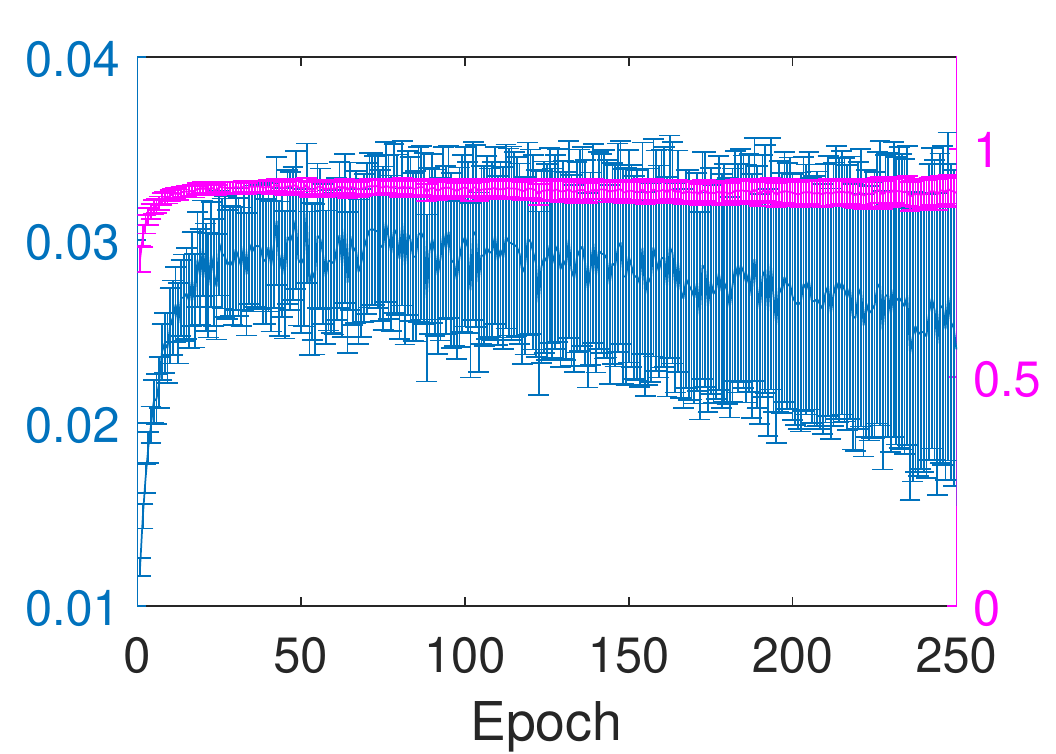}}&
        \hskip -4ex
        \resizebox{\width}{3cm}{\includegraphics[width=0.25\linewidth]{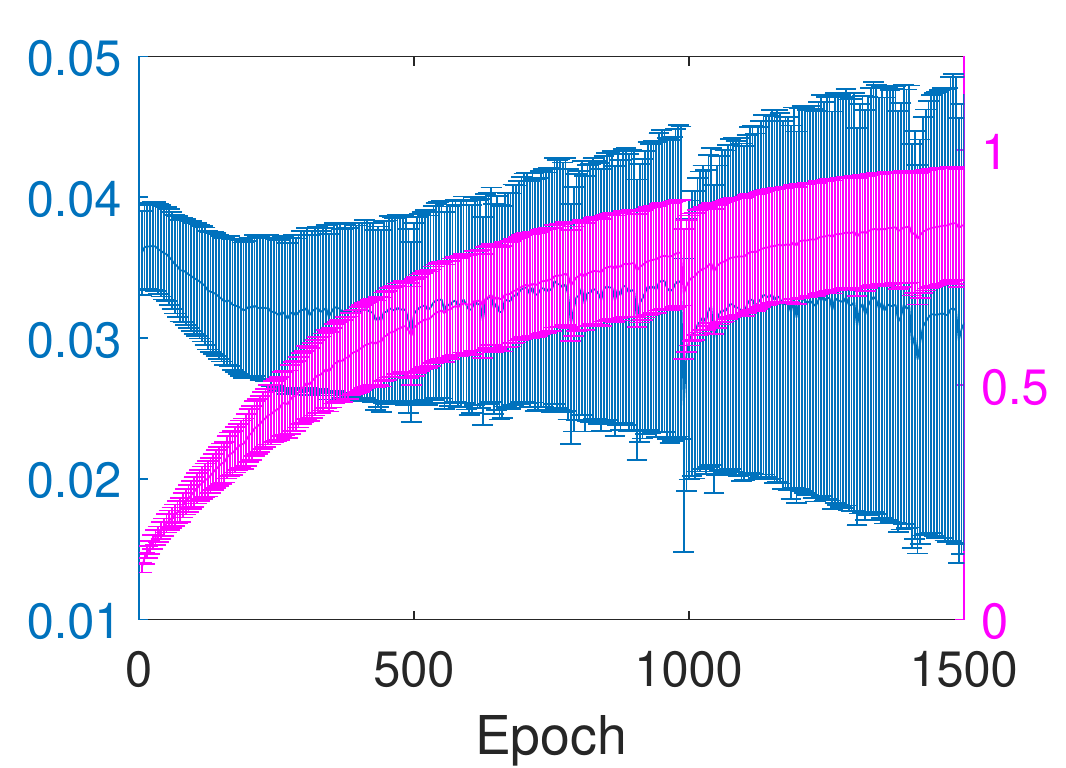}}&
        \hskip -4ex
        \resizebox{\width}{3cm}{\includegraphics[width=0.25\linewidth]{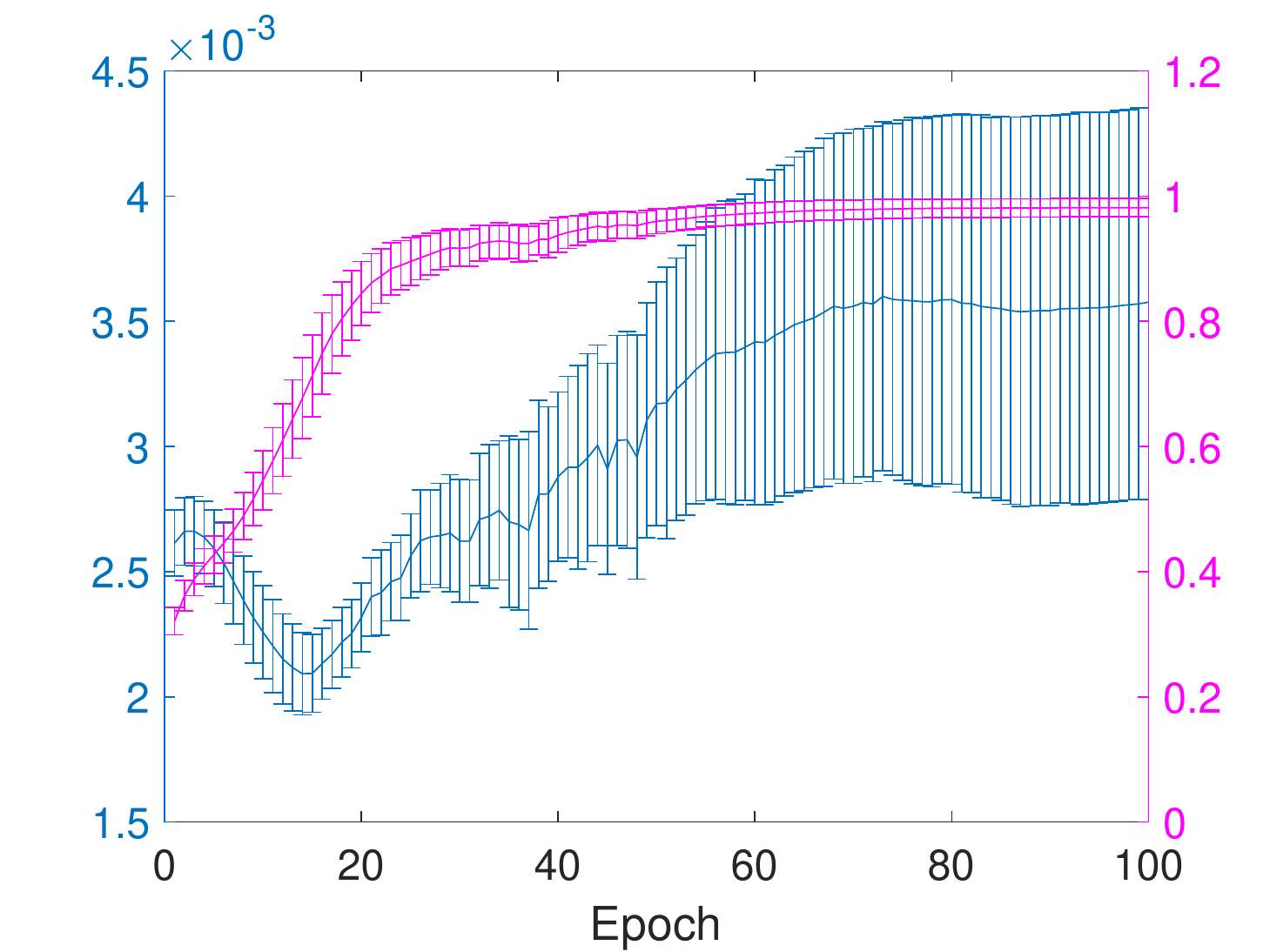}}\\ 
        \begin{sideways}~~~~~~~$L_2,$ zeroed  nodes \end{sideways}&
        \hskip -2ex
        \resizebox{\width}{3cm}{\includegraphics[width=0.25\linewidth]{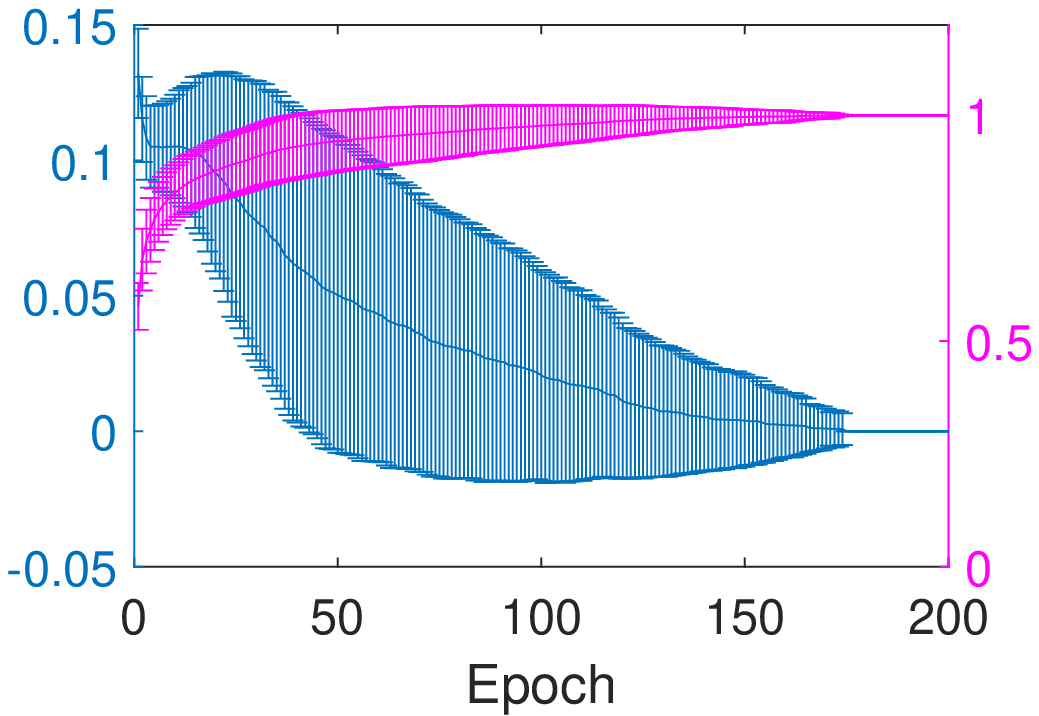}}&
        \hskip -4ex
        \resizebox{\width}{3cm}{\includegraphics[width=0.25\linewidth]{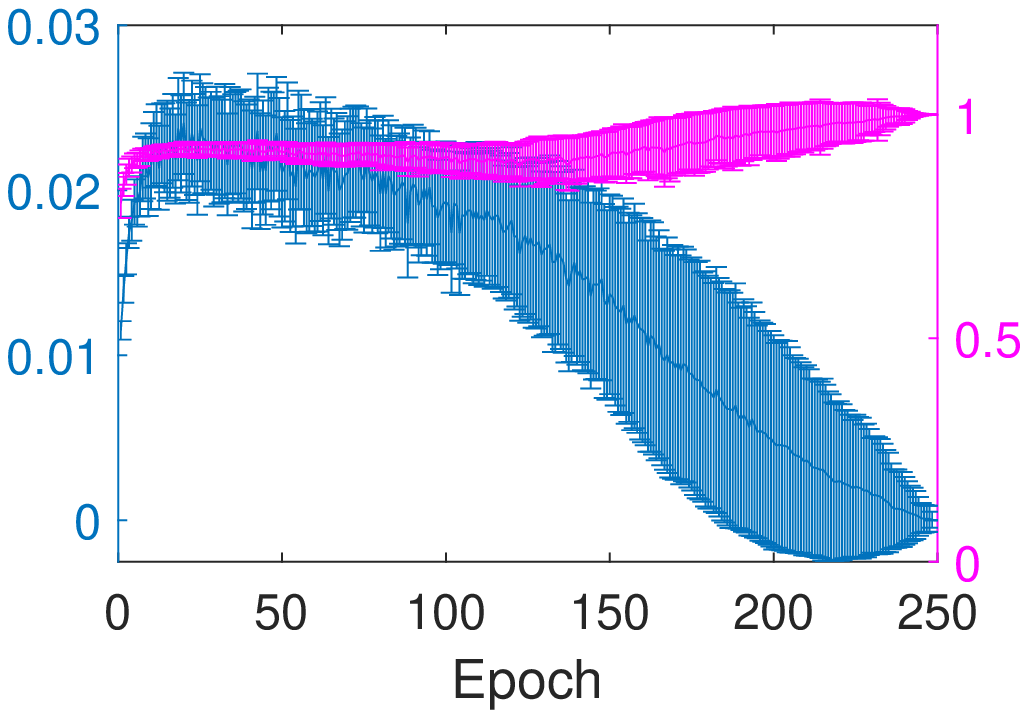}}&
        \hskip -4ex
        \resizebox{\width}{3cm}{\includegraphics[width=0.25\linewidth]{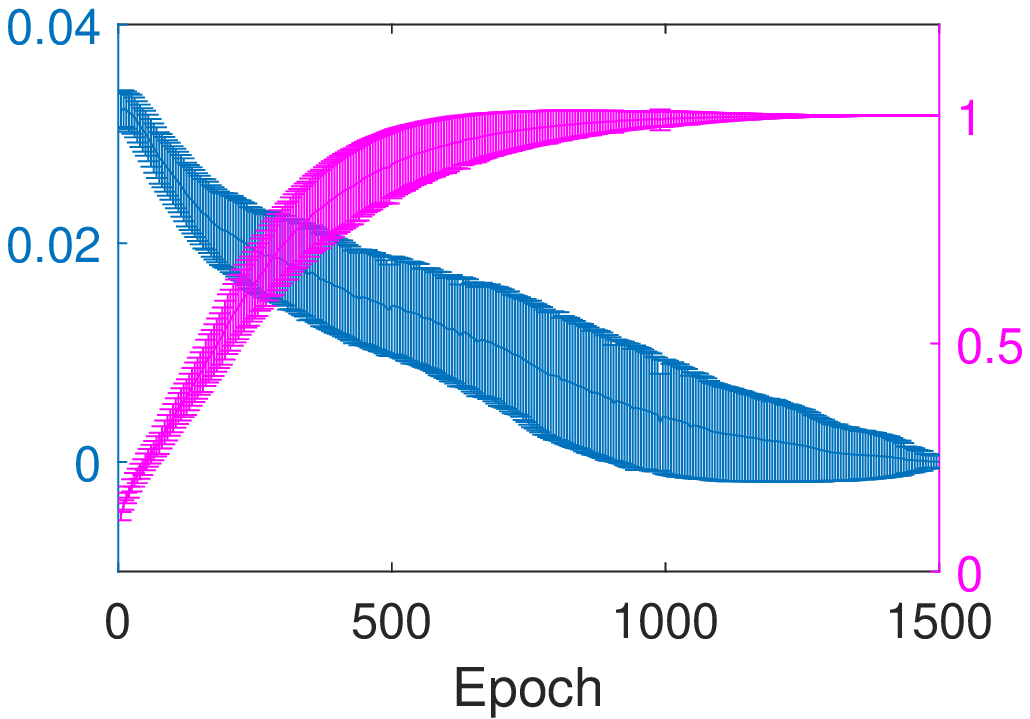}}&
        \hskip -4ex
        \resizebox{\width}{3cm}{\includegraphics[width=0.25\linewidth]{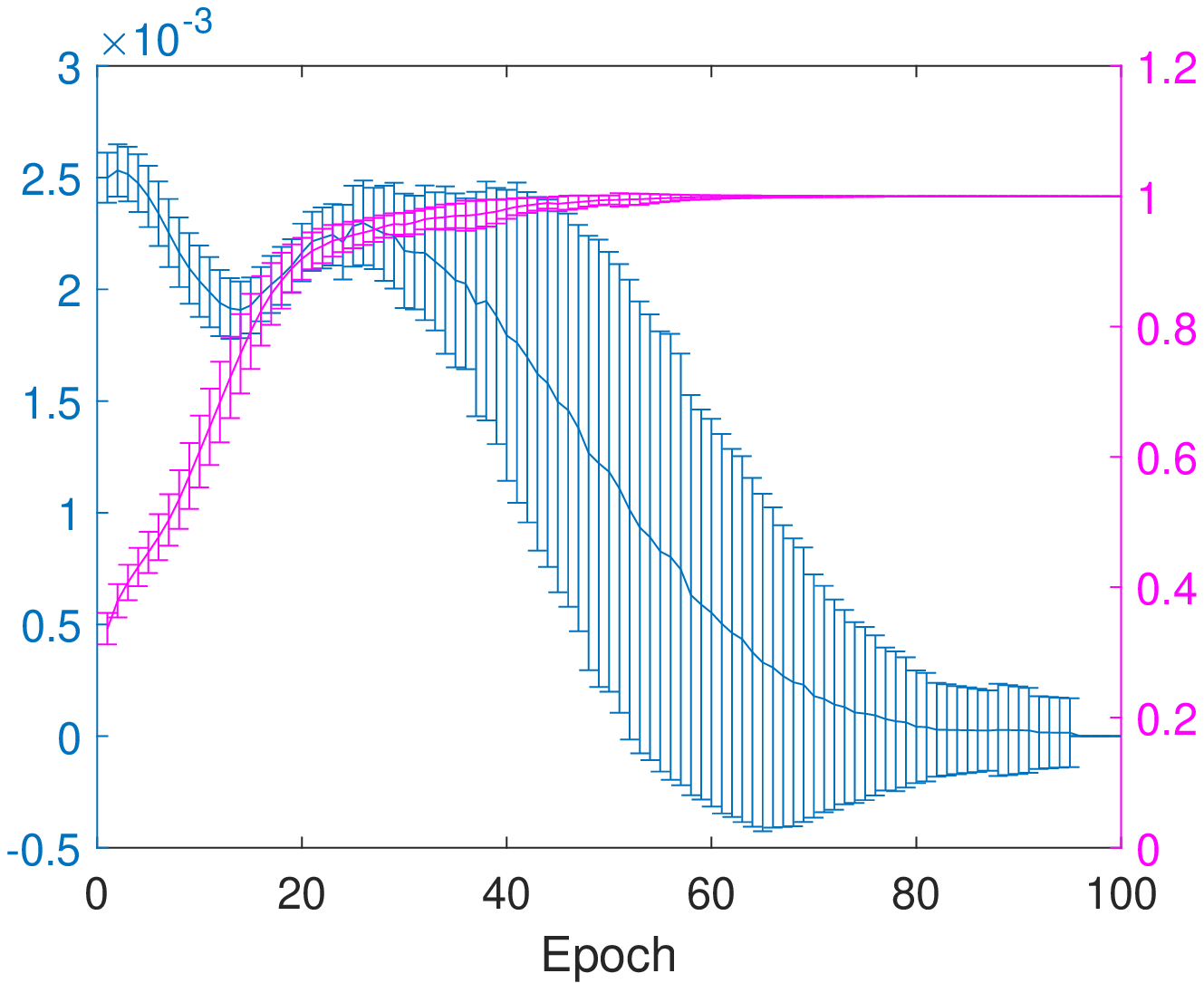}}
        \end{tabular}
    \caption{The mean value of non-zero weights per node (\textcolor{blue}{blue}) and the mean ratio of pruned weights per node (\textcolor{magenta}{magenta}) for each epoch. The first row displays pruning dynamics w/o regularization, the second and third rows display pruning dynamics with $L2$ regularization of the non-zeroed nodes (second) and the zeroed nodes (third).}
    \label{fig:nodesl2Mnist}
\end{figure*}

\section{Analysis of pruning dynamics}
In this section, we investigate the influence of different regularization terms on the dynamics and the structure of weight and node pruning for different network architectures. To highlight the robustness of the pruning dynamics we ran each of the experiments 4 to 6 times for different sets of initial weights. Specifically, we initialized the first training sessions (prior to the pruning) using different sets of random weights. As a result, the initial weights of the  pruning sessions were different for each run. The plots are therefore shown with error bars.\\
{\bf Regularization strength:}
In Figure~\ref{fig:lambda} we present the dynamics of weight and node pruning for the MLP-300-100 using three different values of $L_2$ regularization coefficients $\lambda$. As expected, the percentage of {\bf pruned weights} increased as $\lambda$ took on higher value since the regularization was stronger. However, the change in the percentage of {\bf pruned nodes} as a function of $\lambda$ was higher. 
Table~\ref{tbl:MLPlambda} presents the final pruning results.
\\
{\bf Regularization terms:}
In the next experiment we compared the effects of $L_1,$ $L_2$ and elastic-net regularizations with respect to training with no regularization - on the dynamics of weight and node pruning for four different architectures. The plots presented in Figure~\ref{fig:reg-terms} shows the weight/node pruning percentages as a function of the number of epochs.    
In the absence of a regularization term (purple plots) node pruning percentage was either low (MLP) or negligible (VGG, U-Net, DarkCovidNet) whereas weight pruning percentage was relatively high.  
The plots presented for the MLP-300-100 show that $L_1$ (blue), $L_2$ (red) and the elastic-net (yellow) regularization terms had similar influence on node and weight pruning dynamic and percentages. Different patterns of node pruning dynamics of the VGG and the U-Net architectures were observed for the different regularization terms. It appeared that $L_1$ regularization was preferable for VGG node pruning, yet for weight pruning $L_2$ was preferable.   
Overall using $L_2$ regularization for U-Net provided better segmentation performances, yet better node and weight pruning ratios were obtained by either $L_1$ or elastic-net regularizations.

{\bf Weight compensation and weight decay:}
Finally, we tested our hypothesis that when no regularization is applied, the magnitudes of edges that are neighbors of zero edges (connected to the same nodes) increase to compensate for the missing edges while in the presence of a regularization term, this compensation mechanism is suppressed and gradually entire nodes are zeroed. The results of this experiment are presented in Figure~\ref{fig:nodesl2Mnist}. To assess our assumptions, we chose a specific layer for each of the four architectures. We then calculated, for each epoch, the mean ratio of pruned weights per node (magenta) and the mean magnitude of non-zero weights per node (blue). The size of the error bars presents the average variance per node. The first row in Figure~\ref{fig:nodesl2Mnist} presents the results obtained without regularization. Since a very high percentage of the weights were pruned, the mean ratio of pruned weights per node (magenta plot) almost reached one. Yet, the nodes themselves were not zeroed as the magnitudes of the remaining weights gradually increased to compensate for the missing weights. The second and the third rows of Figure~\ref{fig:nodesl2Mnist} present pruning dynamic results when $L_2$ regularization is applied. The plots in the second row present the results for nodes that were not zeroed throughout the entire pruning process. Similar to the experiments with no regularization (first row) the magnitudes of the remaining weights in the none-zero nodes increased to compensate for the zeroed weights. However, the increment in the mean magnitude was much larger and the variance in magnitude was higher, probably to compensate for the large number of zeroed nodes. The plots in the third row of Figure~\ref{fig:nodesl2Mnist} present the mean results for nodes that eventually decreased to zero. For these nodes the regularization suppressed the tendency of the non-zero weights to grow in order to compensate for their neighboring zeroed weights. Eventually, these weights were also zeroed, as shown in the blue plots. Note that the corresponding average ratio of pruned weights per nodes (magenta) increased to $1.$
\section{Conclusions}
We presented a general stochastic approach for neural network pruning which is indifferent to the network architecture, its training regime and its loss function. The method is shown to be effective for concurrent weight and node pruning. This is accomplished by utilizing weight decay regularization to facilitate pruning and understanding its role in manipulating pruning dynamics. Rather than using a fixed pruning criterion, we defined a probability function and a gating mechanism such that weights with lower absolute values were more likely to be removed. Regularization then played a triple role. Beyond its clear advantage in reducing overfitting, it decreased the weights magnitudes thus increasing their pruning probability. Moreover, while optimizing for weight pruning, node pruning was increased as well. We empirically studied the mechanism for this effect. Specifically, we showed that the $L_1$ and $L_2$ regularization term suppressed the `tendency' of non-zero weights to compensate for neighboring pruned edges associated with the same nodes. Finally, we showed that the  pruning results were on a par with both best the weight and node pruning algorithms for different widely used image classification and segmentation architectures.

\bibliographystyle{cas-model2-names}
\bibliography{Neurocomputing}


\end{document}